\documentclass[11pt]{article}

\usepackage{acl}

\usepackage{times}
\usepackage{latexsym}

\usepackage[T1]{fontenc}
\usepackage{amsmath}
\usepackage[most]{tcolorbox}
\usepackage{xcolor}
\usepackage{amssymb, amsmath}
\usepackage{bbm}
\definecolor{thinkbg}{HTML}{D9ECFC}
\definecolor{thinktext}{HTML}{4E95D9}
\definecolor{llmbg}{HTML}{E8D6EC}
\definecolor{llmtext}{HTML}{6F3C7A}
\definecolor{searchbg}{HTML}{DCDCDC}
\definecolor{searchtext}{HTML}{808080}
\definecolor{infobg}{HTML}{F3E4D4}
\definecolor{infotext}{HTML}{9A6B3D}
\definecolor{answerbg}{HTML}{FBE1E1}
\definecolor{answertext}{HTML}{C55A5A}
\newtcbox{\thinktag}{
  on line,
  enhanced,
  frame hidden,
  boxrule=0pt,
  colback=thinkbg,
  coltext=thinktext,
  arc=1.5mm,          % 圆角变小
  left=0.6mm,
  right=0.6mm,
  top=0.5mm,
  bottom=0.5mm,
  boxsep=0pt,
}
\newtcbox{\graphragtag}{
  on line,
  enhanced,
  frame hidden,
  boxrule=0pt,
  colback=green!12!white,
  coltext=green!50!black,
  arc=1.5mm,          % 圆角变小
  left=0.6mm,
  right=0.6mm,
  top=0.5mm,
  bottom=0.5mm,
  boxsep=0pt,
}
\newtcbox{\llmtag}{
  on line,
  enhanced,
  frame hidden,
  boxrule=0pt,
  colback=llmbg,
  coltext=llmtext,
  arc=1.5mm,          % 圆角变小
  left=0.6mm,
  right=0.6mm,
  top=0.5mm,
  bottom=0.5mm,
  boxsep=0pt,
}
\newtcbox{\searchtag}{
  on line,
  enhanced,
  frame hidden,
  boxrule=0pt,
  colback=searchbg,
  coltext=searchtext,
  arc=1.5mm,
  left=0.6mm,
  right=0.6mm,
  top=0.5mm,
  bottom=0.5mm,
  boxsep=0pt,
}
\newtcbox{\infotag}{
  on line,
  enhanced,
  frame hidden,
  boxrule=0pt,
  colback=infobg,
  coltext=infotext,
  arc=1.5mm,          % 圆角变小
  left=0.6mm,
  right=0.6mm,
  top=0.5mm,
  bottom=0.5mm,
  boxsep=0pt,
}
\newtcbox{\answertag}{
  on line,
  enhanced,
  frame hidden,
  boxrule=0pt,
  colback=answerbg,
  coltext=answertext,
  arc=1.5mm,          % 圆角变小
  left=0.6mm,
  right=0.6mm,
  top=0.5mm,
  bottom=0.5mm,
  boxsep=0pt,
}
\usepackage[most]{tcolorbox}
\definecolor{boxcolor}{RGB}{46,139,87}

\tcbset{
  mygreenbox/.style={
    colback=boxcolor!10!white, % 背景色
    colframe=boxcolor!70!black, % 边框色
    boxrule=0pt,          % 边框粗细
    arc=2pt,                % 圆角
    left=6pt, right=6pt, top=6pt, bottom=6pt % 内边距
  }
}
\tcbset{
  mygraybox/.style={
    colback=black!5,      % 浅灰背景
    colframe=black!35,    % 灰色边框
    boxrule=0pt,
    fonttitle=\Large\bfseries,
    fontupper=\large,
    arc=2pt,
    left=6pt, right=6pt, top=6pt, bottom=6pt
  }
}
\tcbset{
  mypurplebox/.style={
    colback=purple!8,     % 浅紫背景
    colframe=purple!60,   % 深紫边框
    boxrule=0pt,
    fonttitle=\Large\bfseries,
    fontupper=\large,
    arc=2pt,
    left=6pt, right=6pt, top=6pt, bottom=6pt
  }
}
\tcbset{
  mybluebox/.style={
    colback=blue!4,                 % 主体背景（很浅）
    colframe=blue!55!black,         % 边框（加一点黑更稳）
    boxrule=0pt,
    fonttitle=\Large\bfseries,
    fontupper=\large,
    arc=2pt,
    left=6pt, right=6pt, top=6pt, bottom=6pt
  }
}

\tcbset{
  casestudybox/.style={ % Chuanhao
    enhanced,
    breakable,
    colback=blue!5,
    colframe=blue!35!black,
    boxrule=0.8pt,
    arc=3mm,
    outer arc=3mm,
    width=0.96\textwidth,
    left=4mm,
    right=4mm,
    top=2mm,
    bottom=2mm,
    colbacktitle=blue!75!black,
    coltitle=white,
    fonttitle=\normalsize\bfseries
  }
}

\usepackage[utf8]{inputenc}

\usepackage{microtype}

\usepackage{inconsolata}

\usepackage{graphicx}
\usepackage{xcolor}
\definecolor{myblue}{RGB}{46,84,134} % Chuanhao
\usepackage{booktabs}   % \toprule \midrule \bottomrule \cmidrule
\usepackage{multirow}   % \multirow
\usepackage{xcolor}     % 如果别处还用到颜色
\usepackage{array}      % 自定义 L{..} C{..} 列格式
\usepackage{subcaption}
\usepackage{pifont}
\usepackage{caption} % Chuanhao
\usepackage{enumitem}
\newcommand{\cmark}{\ding{51}}
\newcommand{\xmark}{\ding{55}}

\newcolumntype{L}[1]{>{\raggedright\arraybackslash}m{#1}}
\newcolumntype{C}[1]{>{\centering\arraybackslash}m{#1}}
\newcolumntype{Y}{>{\raggedright\arraybackslash}X}

\title{GraphRAG-Router: Learning Cost-Efficient Routing over GraphRAGs and LLMs with Reinforcement Learning}

\author{
 \textbf{Dongzhe Fan\textsuperscript{1}},
 \textbf{Chuanhao Ji\textsuperscript{1}},
 \textbf{Zimu Wang\textsuperscript{2}},
 \textbf{Tong Chen\textsuperscript{2}},
 \textbf{Qiaoyu Tan\textsuperscript{1}}
\\
\\
 \textsuperscript{1}Department of Computer Science, New York University (Shanghai),\\
 \textsuperscript{2}Department of Computer Science, University of Liverpool
\\
 \small{
 \{df2362, cj2851, qiaoyu.tan\}@nyu.edu,
 \{zimu.wang, tong.chen\}@liverpool.ac.uk
 }
}

\begin{document}
\maketitle
\begin{abstract}
Graph-based retrieval-augmented generation (GraphRAG) has recently emerged as a powerful paradigm for knowledge-intensive question answering, especially for tasks that require structured evidence organization and multi-hop reasoning. However, existing GraphRAG systems are typically built in a one-size-fits-all manner, relying on a fixed retrieval framework and a single, often large and costly, generator LLM for all queries. This static design limits their ability to adapt to the complexity of varying questions and often incurs unnecessary computational cost. To fill in the gap, we propose GraphRAG-Router, a cost-efficient framework that adopts a hierarchical routing strategy to coordinate heterogeneous GraphRAGs and generator LLMs. Specifically, GraphRAG-Router is first warmed up through supervised fine-tuning and then optimized with a two-stage reinforcement learning procedure, whose second stage introduces a curriculum cost-aware reward to encourage difficulty-aware and economical generator allocation. Extensive experiments on six general-domain and multi-hop QA benchmarks show that GraphRAG-Router consistently outperforms state-of-the-art baselines, reducing the overuse of large LLMs by nearly 30\% while maintaining strong generalization capability.

% Graph-based retrieval-augmented generation (GraphRAG) has recently emerged as a powerful approach for knowledge-intensive question answering, particularly for tasks requiring structured evidence organization and multi-hop reasoning. However, existing GraphRAG systems typically rely on fixed retrieval infrastructure and single large generator LLM, limiting their ability to adapt to varying question complexity and leading to unnecessary computational cost. To address this limitation, we propose GraphRAG-Router, a cost-efficient framework that adopts a hierarchical routing strategy to coordinate heterogeneous GraphRAGs and generator LLMs. Warmed up by supervised fine-tuning , GraphRAG-Router is optimized with a two-stage RL paradigm, in which the second stage introduces a curriculum cost-aware reward to encourage difficulty-aware and cost-efficient generator allocation. Extensive experiments on six general and multihop QA benchmarks demonstrate that GraphRAG-Router
% consistently surpasses state-of-the-art baselines, reduces the overuse of large LLMs by nearly $30\%$ while maintaining strong generalization capability. Our code is anonymously shared at \url{https://anonymous.4open.science/r/GraphRAG-Router-101}
\end{abstract}

\section{Introduction}
%\dz{Paragraph1:Graph-based retrieval-augmented generation (GraphRAG) has emerged as a promising paradigm for knowledge-intensive QA, as it can organize evidence into structured entities and relations and thus better support multi-hop reasoning. Recently...Graph-R1/Search-R1... However, }\\
Graph-based retrieval-augmented generation (GraphRAG) has emerged as a powerful paradigm for knowledge-intensive question answering (QA)~\cite{G-retriever,RAPTOR,GraphRAG,HippoRAG}, especially for multi-hop questions that require structured evidence organization and compositional reasoning. By organizing knowledge into entities, relations, and higher-level graph structure, GraphRAG enables more explicit evidence aggregation than standard text-based retrieval-augmented generation, making it particularly well suited for complex reasoning over dispersed evidence~\cite{han2024retrieval,peng2025graph}.
% Recent GraphRAG methods have shown strong promise across a variety of QA settings, demonstrating the value of structured retrieval for improving reasoning quality and interpretability.
\begin{figure}[t]
    \centering
    \begin{subfigure}[t]{0.49\linewidth}
        \centering
        \includegraphics[width=\linewidth]{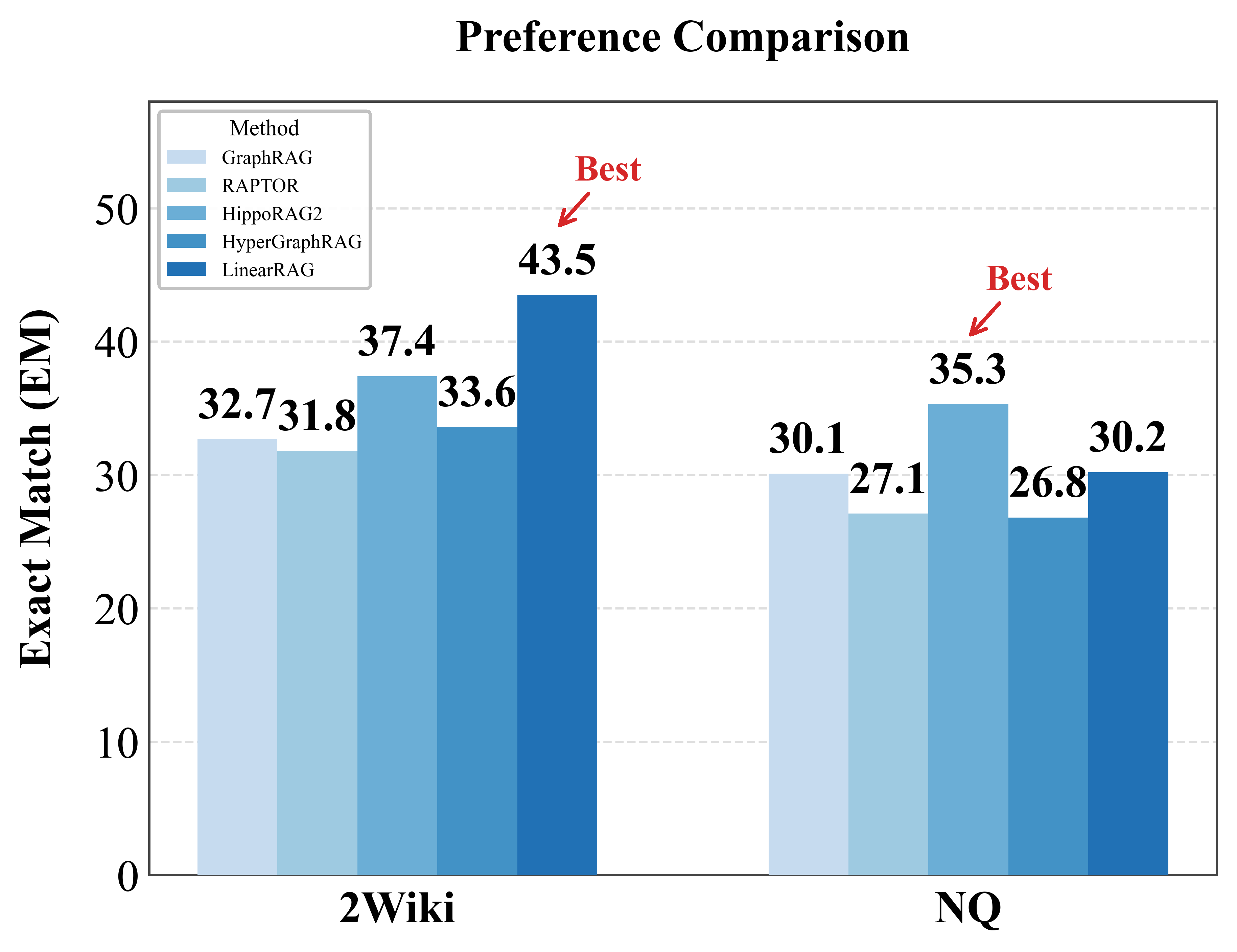}
        \caption{Dataset preference of GraphRAGs.}
        \label{fig:motivation1}
    \end{subfigure}
    \hfill
    \begin{subfigure}[t]{0.49\linewidth}
        \centering
        \includegraphics[width=\linewidth]{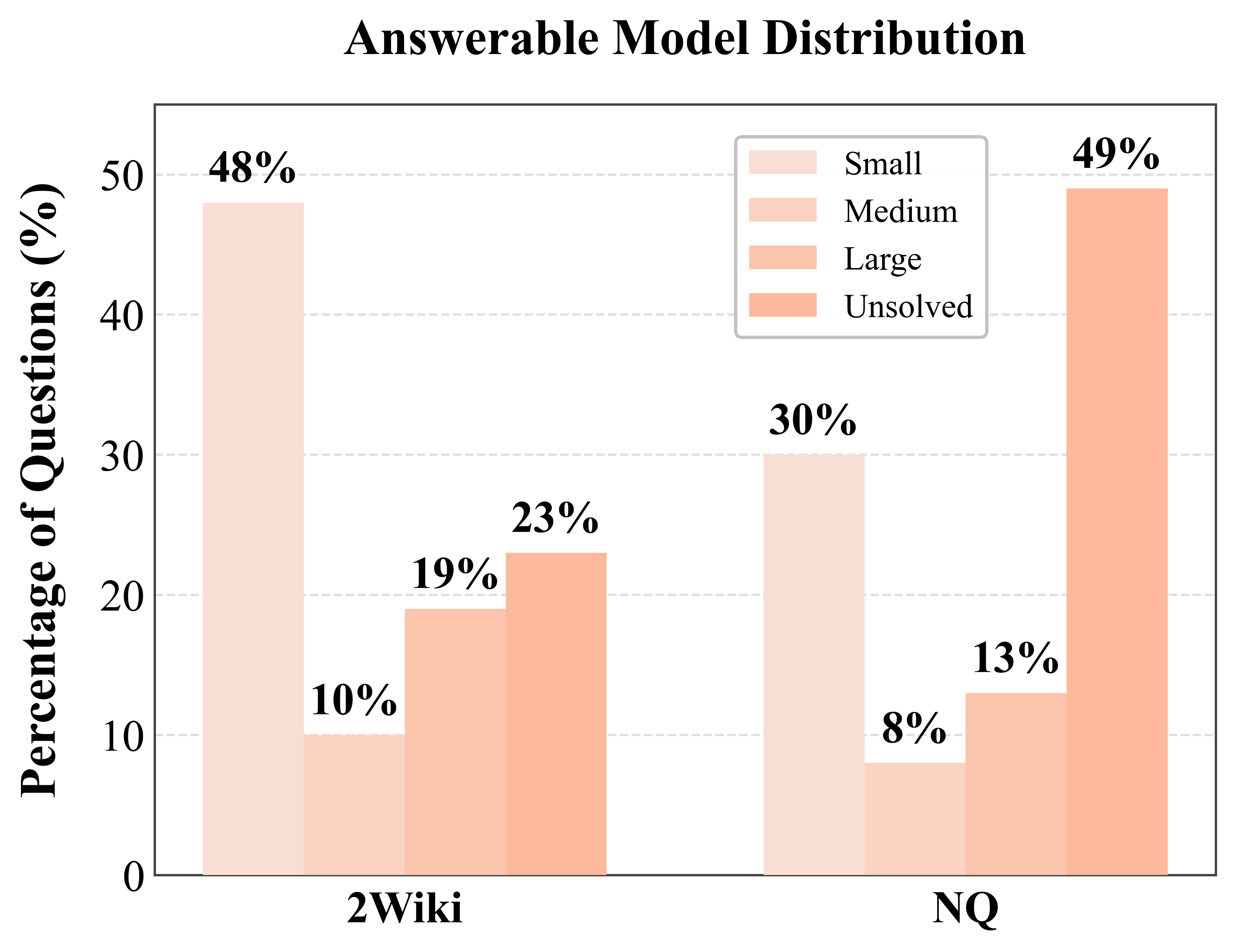}
        \caption{Distribution of answerable model scales.}
        \label{fig:motivation2}
    \end{subfigure}
    \caption{Query-dependent heterogeneity in GraphRAGs and generator LLMs.}
    \label{fig:main_figure}
\vspace{-10pt}
\end{figure}

Despite these advances, current GraphRAG systems are still largely built in a \textit{one-size-fits-all manner}: they typically rely on a single retrieval framework and a single generator LLM for all queries. Such a static design is inherently suboptimal, because questions vary substantially in both the form of evidence they require and the level of reasoning capacity needed to answer them. On the retrieval side, \textbf{different QA benchmarks exhibit markedly different preferences over GraphRAG frameworks} (Figure~\ref{fig:motivation1}), suggesting that different questions demand evidence at different levels of granularity. For instance, entity-centric lookup questions are better served by local subgraph retrieval, whereas questions requiring thematic synthesis benefit more from community-level summarization. On the generation side, \textbf{existing GraphRAG systems often rely on a large and costly LLM to ensure strong overall performance}~\cite{GraphRAG, HippoRAG2} (e.g., GPT-4o~\cite{GPT4}), yet Figure~\ref{fig:motivation2} shows that many questions can already be solved correctly by much smaller models. Uniformly invoking a large model therefore wastes computation, while relying only on a smaller model may fail on more challenging queries. These observations indicate that both retrieval infrastructure and generator allocation should be adaptive, so that each query can be matched with the most appropriate evidence source and model capacity. This raises a central question:
\textit{How can we adaptively coordinate multiple GraphRAGs and generator LLMs for each query, while maintaining strong QA performance without incurring unnecessary generator cost?}

Solving this problem is non-trivial due to two major challenges. (1) \textbf{Combinatorial action space in joint GraphRAG--LLM selection.} While prior studies have explored model-wise routing among candidate LLMs~\cite{Router-R1,KNNRouter}, our setting requires the routing agent to jointly select both the GraphRAG  and the generator LLM at each reasoning step. This joint decision substantially expands the decision space compared to model-only routing, and the exponential growth of the action space makes effective optimization particularly challenging.
(2) \textbf{Cost-efficient routing.}
An effective routing strategy must balance answer quality against the computational overhead of generator LLM invocation.
Since queries differ widely in their reasoning complexity, the optimal model scale, defined as the minimum capacity required to produce a correct response, is inherently context-dependent. Therefore, the routing policy must calibrate its allocations precisely, as over-provisioning leads to unnecessary computational cost, while under-provisioning risks degrading answer accuracy.
%Although prior routing methods~\cite{} attempt to solve queries with different LLMs, their routing distributions still tend to concentrate on large-scale models, leading to substantial inference costs. However, as illustrated in Figure~\ref{fig:motivation2}, a large portion of questions can in fact be handled effectively by small-scale LLMs. suggesting that existing approaches do not sufficiently exploit cheaper models when they are adequate. Therefore, beyond correctness, the router must also learn to calibrate model capacity to query difficulty, so as to achieve a better trade-off between performance and inference cost.

To this end, we introduce \textbf{GraphRAG-Router}, a cost-efficient RL framework that enables multi-round joint routing and aggregation across heterogeneous GraphRAG frameworks and generator LLMs. Instead of making a single monolithic dispatch decision over an LLM--GraphRAG pair, we formulate the coordination process as a \textit{hierarchical routing path} that allows the model to select the GraphRAG framework and generator LLM step by step.
Specifically, we first warm up the routing policy via supervised fine-tuning (SFT) using curated routing trajectories. We then optimize the policy with a two-stage RL procedure that incorporates three complementary reward signals: a \textit{format reward} encouraging well-structured outputs, an \textit{outcome reward} based on final task correctness, and a \textit{curriculum cost-aware reward} that promotes difficulty-aware model utilization by penalizing unnecessarily expensive routing decisions. Together, these components make GraphRAG-Router a cost-efficient and generalizable solution for multi-round coordination across heterogeneous GraphRAG frameworks and LLMs, achieving state-of-the-art performance on diverse knowledge-intensive QA benchmarks.

The key contributions of this work can be summarized as follows:

\begin{itemize}[itemsep=0pt,topsep=0pt,leftmargin=*,label=$\star$]
    \item We introduce \textbf{GraphRAG-Router}, an RL--based framework for multi-round routing and aggregation across heterogeneous GraphRAGs and generator LLMs, enabling question-adaptive routing across retrieval and generation modules.
    \item We propose a hierarchical routing strategy and optimize GraphRAG-Router with a two-stage RL procedure with a curriculum cost-aware reward, allowing the model to route queries to appropriate GraphRAG--LLM pairs while better balancing performance and LLM  cost.
    \item Experimental results on six QA benchmarks demonstrate that GraphRAG-Router consistently outperforms strong baselines, achieving superior cost-efficiency, robust generalization, and state-of-the-art overall performance.
\end{itemize}

%\dz{Paragraph4: Our contribution}
\begin{figure*}[t]
\centering
\includegraphics[width=15cm, height=6.5cm]{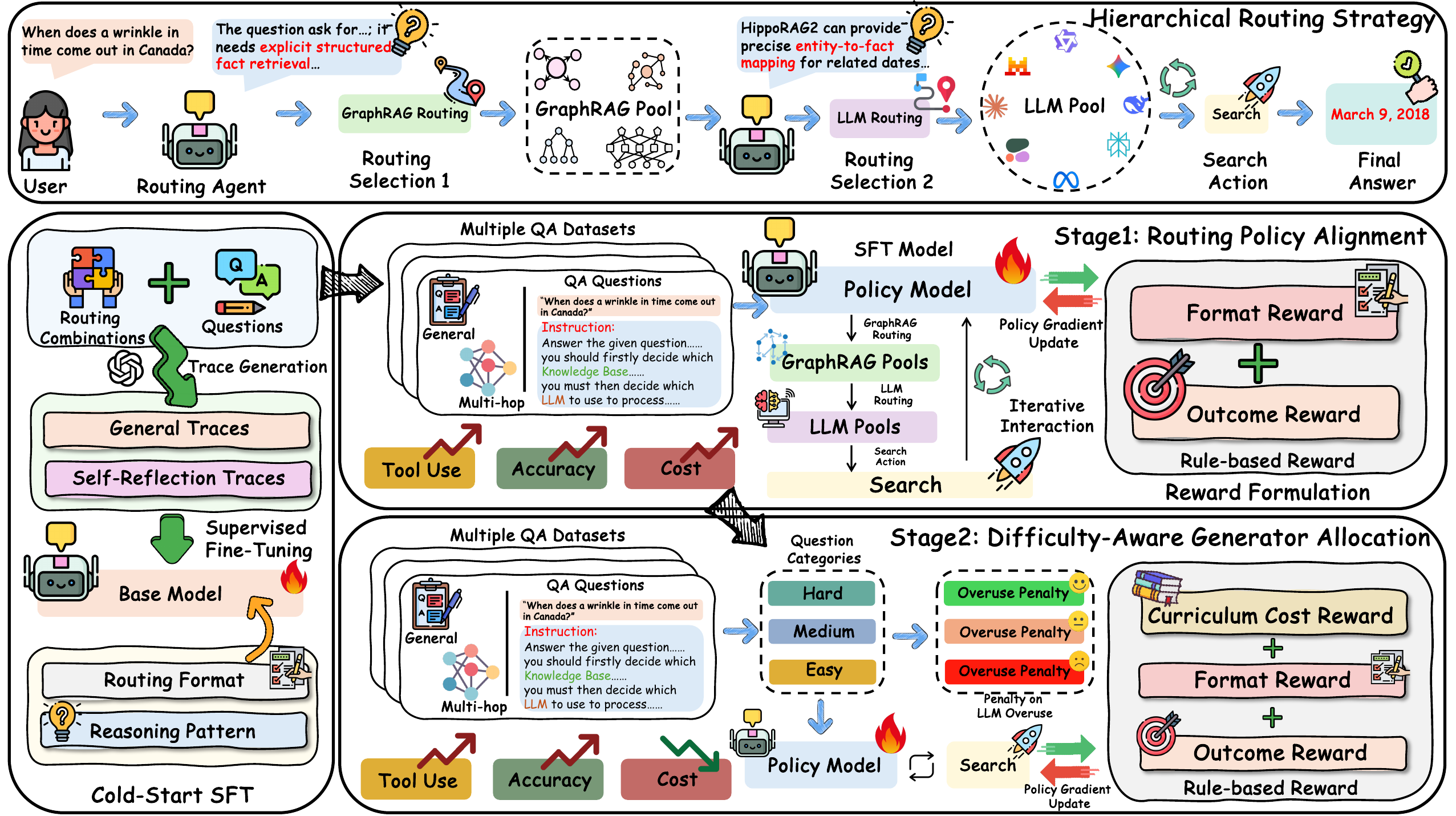}
\caption{\textbf{Overview of the GraphRAG-Router framework. GraphRAG-Router utilizes a hierarchical routing strategy. Initialized with cold-start SFT, it adopts a two-stage RL training paradigm. Stage 1 helps the model learn effective routing behavior, while Stage 2 further optimizes the trade-off between performance and cost.
}}
\label{fig:main}
\vspace{-10pt}
\end{figure*}

\section{Related Work}
\paragraph{GraphRAG.}
GraphRAG extends the conventional RAG paradigm by structuring retrieved knowledge as entities and relations, thereby providing stronger support for multi-hop reasoning and knowledge-intensive QA. Early GraphRAG methods~\cite{GraphRAG,RAPTOR} typically construct knowledge graphs with entities as nodes and relations as edges. Recent works~\cite{LightRAG,LinearRAG,HippoRAG,HippoRAG2} further enhance the representation power of GraphRAG by incorporating richer graph structures (e.g., HyperGraph), thereby improving evidence organization and multi-hop reasoning over complex knowledge, typically with the support of large-scale LLMs. Meanwhile, another line of work explores GraphRAG systems~\cite{Graph-R1, GraphSearch, GraphRAG-R1} that rely on smaller models, whose reasoning capabilities are enhanced through RL. In this study, we focus on the setting where GraphRAG frameworks are paired with large and costly LLMs, which often incurs substantial computational cost and still relies on a fixed retrieval infrastructure.

\paragraph{LLM Routing.}
Current LLM routing frameworks aim to select appropriate generator models from a candidate pool according to query complexity. Existing studies have explored both predictive model routing~\cite{KNNRouter, graphrouter} and multi-round routing strategies over candidate LLMs, including RL-based approaches such as Router-R1~\cite{Router-R1} that interleave internal reasoning with dynamic model selection and response aggregation. However, these methods focus solely on routing among generator LLMs. In contrast, our setting requires joint routing over both GraphRAG frameworks and generator LLMs, which introduces a substantially larger and more complex action space.

% \paragraph{Agentic Search LLMs.} To mitigate the limitations of single-shot retrieval, recent works~\cite{Search-o1,GraphSearch,GraphSearchjiajin} has explored agentic paradigms in which LLMs iteratively conduct multi-round search and reasoning, progressively collecting evidence for complex tasks. More recently, with the emergence of reinforcement learning methods such as GRPO~\cite{GRPO}, some studies~\cite{Graph-R1,Search-R1,GraphRAG-R1} have moved beyond training-free designs and explored learnable search policies, where multi-turn search behavior is optimized with rule-based reward signals.

\section{Problem Definition}
We formulate our task as follows: Given a query \(q\), a candidate pool of GraphRAGs
\(\mathcal{G}=\{g_1,g_2,\dots,g_n\}\),
and a candidate pool of generator LLMs
\(\mathcal{L}=\{l_1,l_2,\dots,l_m\}\),
the goal is to learn a routing policy \(\pi_{\theta}\) that selects GraphRAGs and generator LLMs over multiple rounds to produce the final answer \(a\) under the instruction \(\mathcal{T}\).
At each round \(t\), the router first performs internal reasoning \(\mathcal{R}_t\) to select a GraphRAG \(g_t \in \mathcal{G}\) and a generator LLM \(l_t \in \mathcal{L}\) based on the current reasoning state \(s_t\). The selected GraphRAG--LLM pair then returns information $I_t = l_m\bigr(q, g_t(q)\bigl)$. 
The information is incorporated into the state for subsequent states, progressively constructing the final answer:
\[
\small
\begin{aligned}
& P(\mathcal{R}, a \mid q, \mathcal{T}, I)= \\
&\underbrace{\prod_{t=1}^{T_r}
P(\mathcal{R}_t \mid \mathcal{R}_{<t}, q, \mathcal{T},I_{<t})}_{\text{Reasoning Process}}
\cdot \underbrace{\prod_{t=1}^{T_a}
P(a_t \mid a_{<t}, \mathcal{R}, q, \mathcal{T})}_{\text{Answer Generation}}
\end{aligned}
\]
Where $T_r$ and $T_a$ are the number of tokens in the reasoning sequence and answer, respectively. $I_{<t}$ denotes all the returned information up to token $t$.

% \noindent \textbf{Routing Process.}
% At each round $t$, the router first performs internal reasoning based on the previous steps to generate a sub-query $q_t$, and then selects a GraphRAG framework $g_t$ and a generator LLM $l_t$ accordingly. The selected GraphRAG framework $g_t$ is then invoked to retrieve evidence conditioned on $q_t$.
% \[
% E_t = g_t(q_t),
% \]
% The retrieved evidence is then passed to the generator LLM to produce the next-step information:
% \[
% I_t = l_t(q_t, E_t),
% \]
% where $l_t \in \mathcal{L}$ is the selected generator LLM, and $I_t$ denotes its output. The output $I_t$ can either directly yield the answer $a_t$ or provide information that guides the router’s next action in subsequent rounds.

% \noindent \textbf{Answer Generation.}
% After multi-round interaction, the router yields the final answer $a$, grounded in previous routing process:
% \[
% \begin{aligned}
% & P(\mathcal{R}, a \mid q, \mathcal{G}, \mathcal{L})= \\
% &\underbrace{\prod_{t=1}^{T_r}
% P(\mathcal{R}_t \mid \mathcal{R}_{<t}, q, I_{<t})}_{\text{Routing Process}}
% \cdot \underbrace{\prod_{t=1}^{T_a}
% P(a_t \mid a_{<t}, \mathcal{R}, q, I)}_{\text{Answer Generation}} .
% \end{aligned}
% \]
% where, $T_r$ is the number of tokens in the reasoning sequences $\mathcal{R}$, $I_{t}$ represents all the information from generator LLMs up to token $t$ in the reasoning chain. Similarly, $T_a$ is the length of the answer sequence $a$.

\section{Methods}
In this section, we detail GraphRAG-Router into four components: Section~\ref{sec:HRP} describes the hierarchical routing strategy over candidate GraphRAGs and generator LLMs. Section~\ref{sec:SFT} introduces the SFT paradigm, which equips the model with well-structured reasoning and routing formats. Sections~\ref{sec:stage1} and \ref{sec:stage2} describe two complementary RL-based training stages: Stage 1 enables the model to acquire core routing behavior, while Stage 2 further regulates the invocation of generator LLMs to achieve cost-efficient routing.
\subsection{Hierarchical Routing Strategy}
\label{sec:HRP}
To make GraphRAG--LLM coordination tractable, we formulate GraphRAG-Router as a hierarchical routing strategy that progressively selects retrieval infrastructures and generators through reasoning. Upon receiving a question, the model first performs internal analysis within \thinktag{<think>} and \thinktag{</think>} to assess the query’s evidence requirement, and then selects a GraphRAG within \graphragtag{<graphrag>} and \graphragtag{</graphrag>} based on the predefined GraphRAG pool and its descriptions, such as evidence granularity and retrieval specialization. The model then performs further reasoning based on the selected GraphRAG, the query, and the expected type and granularity of the retrieved evidence, and selects the most suitable generation LLM within \llmtag{<llm>} and \llmtag{</llm>} accordingly. Subsequently, it queries the specialist generation LLM with retrieved evidence via \searchtag{<search>}\textit{Query:LLM;GraphRAG} \searchtag{</search>}. The resulting information is returned within \infotag{<information>} and \infotag{</information>} tags. The model may iteratively route across multiple GraphRAG--LLM pairs to gather complementary insights, which are then integrated to produce the final answer within \answertag{<answer>} and \answertag{</answer>}. This hierarchical routing strategy reduces the per-step action space from $\mathcal{O}(|\mathcal{G}|\cdot|\mathcal{L}|)$ joint choices to $\mathcal{O}(|\mathcal{G}|+|\mathcal{L}|)$ staged choices, thereby facilitating more effective optimization.

\subsection{Cold Start Supervised Fine-Tuning}
To equip the model with a proper routing skeleton, long-horizon reasoning ability, and initial routing capability, we first conduct supervised fine-tuning (SFT) on curated routing trajectories. To this end, we construct two types of supervised data, namely general routing traces and self-reflection routing traces, and then train the model on these trajectories to learn well-structured reasoning and routing behaviors.
\label{sec:SFT}

\paragraph{General Routing Trace Generation.} We generate high-quality single-turn trajectories that correctly solve the given question while adhering to the desired reasoning, routing, and tool-use manner. Specifically, we begin by collecting GraphRAG--LLM pairs $(g_t, l_t)$ that can correctly solve the question via direct inference, which serve as the basis for constructing general routing trajectories. We then leverage a strong reasoning model (e.g., GPT-5.2~\cite{GPT5}) to generate complete trajectories, including the rationale for each routing decision, such as GraphRAG and generation LLM selection. More details are provided in Appendix~\ref{app:gen}
%\textcolor{red}{No details, such as prompt template. Refer to the details in Appendix. }

\paragraph{Self-Reflection Routing Trace Generation.}
Since a single routing decision may fail to select the optimal GraphRAG--LLM pair, especially for challenging questions, we further introduce multi-round self-reflective routing traces. To equip the model with self-reflection abilities, we construct self-reflective trajectories on top of single-turn traces. For each question, we first collect multiple GraphRAG–LLM pairs via direct inference. We then use a strong reasoning model to synthesize multi-turn trajectories in which the model reflects on why the previous route failed to solve the question, identifies whether the failure stems from inadequate retrieval or insufficient reasoning ability, and accordingly decides whether to switch to a different GraphRAG or a stronger LLM. In this way, the model learns to diagnose failure, adjust routing decisions, and progressively reach the final answer. More details are provided in Appendix~\ref{app:gen}

\paragraph{Training Paradigm.}
By constructing both general and self-reflection routing trajectories, we collect a high-quality hierarchical routing dataset, denote as $\mathcal{D}_{\mathrm{SFT}}$. We then apply SFT to train the model to conduct expect routing strategy:
\[
\small
\mathcal{L}_{\mathrm{SFT}}
=
-\mathbb{E}_{(x,y)\sim \mathcal{D}_{\mathrm{SFT}}}
\left[
\sum_{t=1}^{|y|}
\log P_\theta(y_t \mid x, y_{<t})
\right]
\]
where $y$ is the synthetic trajectory.

\subsection{Routing Policy Alignment}
\label{sec:stage1}
Following the cold-start phase, which helps the model acquire an initial structured routing pattern and basic tool-use capability, we leverage RL-based training to further enhance the model's tool-use ability and optimize routing strategies. We extend the general RL optimization objective with external routing pools:
\[
\small
\begin{aligned}
\max_{\pi_\theta}\;&\mathbb{E}_{x\sim\mathcal{D},\,y\sim\pi_\theta(\cdot\mid x;\mathcal{G},\mathcal{L})}
\Bigl[r_\phi(x,y) \\
&-\beta \,\mathbb{D}_{\mathrm{KL}}\!\left[
\pi_\theta(y\mid x;\mathcal{G},\mathcal{L}) \,\middle\|\, \pi_{\mathrm{ref}}(y\mid x;\mathcal{G},\mathcal{L})
\right]\Bigr]
\end{aligned}
\]
where $\pi_\theta$ and $\pi_{\mathrm{ref}}$ represent the policy and reference model, respectively, both of which are initialized from the SFT model, $r_\phi$ is the reward function and $\mathbb{D}_{\mathrm{KL}}$ is the KL-divergence. To optimize the hierarchical routing policy, we adopt a rule-based reward function including fine-grained format rewards and final outcome rewards:
\[
r_{\phi(x,y)} = r_{\mathrm{format}(y)} + r_{\mathrm{outcome}}
\]
\paragraph{Fine-grained Format Reward.} Since our framework produces structured hierarchical routing strategy, format correctness is crucial for ensuring valid reasoning traces and executable routing actions. We therefore design a fine-grained format reward that assigns progressively larger penalties to structural errors of different severity, rather than using a binary format signal:
\[
\small
r_{\mathrm{format}}(y)
=
-\min\left(1,\sum_{k=1}^{K}\lambda_k \,\mathbb{I}_k(y)\right)
\]
where $\mathbb{I}_k(y)$ indicates whether the $k$-th format rule is violated. The detailed rules are provided in Appendix~\ref{app:format}. This formulation provides more informative supervision for learning stable multi-round hierarchical routing behavior.\\
\paragraph{Final Outcome Reward.}
In GraphRAG-Router, we leverage Exact Match (EM) to assess the correctness of the final answer generated by the routing agent with respect to the ground truth, and use it as the sole outcome reward to guide optimization:
\[
r_{\mathrm{outcome}} = \mathrm{EM}(a, a_{gt})
\]
\subsection{Difficulty-Aware Generator Allocation}
\label{sec:stage2}
Although the Routing Policy Alignment stage equips the model with foundational routing ability, the learned policy is often feasible but not optimal, since it remains biased toward large-scale generator LLMs. In practice, a substantial fraction of questions can be handled by smaller models, and thus the unnecessary invocation of large-scale models leads to avoidable computational cost. To bridge this gap, we introduce Difficulty-Aware Generator Allocation, which adaptively penalizes the overuse of large-scale generator models according to question difficulty, encouraging cost-efficient routing without compromising answer correctness. Specifically, we first categorize questions by difficulty level, and then design a curriculum cost-aware reward that penalizes unnecessarily expensive model usage accordingly.

\paragraph{Question Difficulty Categorization.} Inspired by \citet{curriculumreinforcementlearningeasy}, we extend the definition of question difficulty to the setting of multi-scale LLM routing. Specifically, we define it as the minimum generator model scale required to answer a question correctly and reliably. Specifically, for each question $q$, we perform direct inference using every generator LLM in the routing pool for $N$ independent trials, and compute the success rate of each model $l_m$ as
\[
\small
\mathrm{SR}(l_m, q)=\frac{c(l_m,q)}{N}
\]
where $c(l_m,q)$ denotes the number of successful trials of model $l_m$ on question $q$. We then regard a model $\mathcal{L}_m$ as being able to reliably solve question $q$ if
\[
\small
\mathrm{SR}(l_m, q)\ge \tau
\]
where $\tau$ is a predefined success-rate threshold. Further, we define the difficulty of question $q$ by the minimum model scale that satisfies this condition:
\[
\small
\begin{aligned}
m(q)=\min \Bigl\{ s \in \{\text{small},\text{medium},\text{large}\} \\
\mid\ \mathrm{SR}(q,s)\ge \tau
\Bigr\}
\end{aligned}
\]
Accordingly, the difficulty level of question $q$ is defined as
\[
\small
\mathrm{D}(q)=
\begin{cases}
\text{Easy}, & m(q)=\text{small}\\
\text{Medium}, & m(q)=\text{medium}\\
\text{Hard}, & m(q)=\text{large}
\end{cases}
\]
\paragraph{Curriculum Cost-aware Reward.}
To facilitate cost-efficient routing, we first assign each generator LLM a predefined cost $C(l_m)$ according to its model scale, so that model usage becomes explicitly measurable. Combined with the above difficulty categorization, this allows us to define, for each question $q$, its minimum required cost $C_{\min}(q)$ as the cost of the least expensive generator LLM that can reliably answer it correctly:
\[
\small
C_{\min}(q)=\min \left\{ C(l_m)\ \middle|\ \mathrm{SR}(l_m,q)\ge \tau \right\}
\]
Based on the minimum required cost $C_{\min}(q)$, we can quantify whether the selected generator LLM is unnecessarily expensive for question $q$. However, such overuse should not be penalized uniformly, since harder questions require greater flexibility to explore stronger models. We therefore adopt a curriculum design that adjusts the penalty strength according to question difficulty:
\[
\small
r_{\mathrm{cost}}(y)=
\beta\, w_{\mathrm{D}(q)} \, \max \bigl(0,\, C(l_m) - C_{\min}(q)\bigr)
\]
where $C(l_m)$ denotes the cost of the generator LLM selected for question $q$, $\beta$ is a scaling coefficient, and $w_{\mathrm{Difficulty}(q)}$ is the penalty weight associated with the difficulty level of $q$, with larger weights assigned to easier questions and smaller weights assigned to harder ones.

\paragraph{Reward Shaping.} To sum up, the overall reward in the Difficulty-Aware Cost Optimization stage is formulated as:
\[
r_{\phi(x,y)} = r_{\mathrm{format}(y)} + r_{\mathrm{outcome}} - \mathbbm{1}\{\mathrm{Correct}\}\, r_{\mathrm{cost}(y)}
\]
where $\mathbbm{1}\{\mathrm{\mathrm{Correct}}\}$ indicates whether the trajectory yields the correct answer. This design encourages the model to improve cost efficiency under successful task completion, naturally converging to the minimal sufficient routing trajectory.

\begin{table*}[t]
\centering
\small
\setlength{\tabcolsep}{2pt}
\renewcommand{\arraystretch}{1.10}
\label{tab:1}

\begin{tabular}{@{} L{2.8cm} L{3.2cm} *{7}{C{1.35cm}} @{}}
\toprule
\multirow{2}{*}{\textbf{System}} &
\multirow{2}{*}{\textbf{Methods}} &
\multicolumn{3}{c}{\textbf{General QA}} &
\multicolumn{4}{c}{\textbf{Multi-Hop QA}} \\
\cmidrule(r){3-5}\cmidrule(l){6-9}
& & \textbf{NQ$^\dagger$} & \textbf{PopQA$^\star$} & \textbf{TriviaQA$^\star$}
  & \textbf{HotpotQA$^\dagger$} & \textbf{2Wiki$^\star$} & \textbf{Musique$^\star$} & \textbf{Avg.} \\
\midrule

\multirow{2}{*}{\textbf{Basic LLM}} &
Direct Infer$^\spadesuit$ & 0.107 & 0.117 & 0.308 & 0.269 & 0.295 & 0.118 & 0.202 \\
& CoT$^\spadesuit$ & 0.286 & 0.327 & 0.552 & 0.313 & 0.293 & 0.165 & 0.323 \\
\midrule

\multirow{6}{*}{\textbf{RAG-Based}} &
Vanilla RAG & 0.294 & 0.242 & 0.557 & 0.326 & 0.348 & 0.186 & 0.326 \\
& HippoRAG2 & 0.353 & 0.266 & 0.601 & \underline{0.395} & 0.374 & 0.263 & 0.375 \\
& LinearRAG & 0.302 & 0.269 & 0.608 & 0.385 & 0.435 & 0.238 & 0.373 \\
& HyperGraphRAG & 0.268 & 0.272 & 0.585 & 0.253 & 0.336 & 0.209 & 0.321 \\
& RAPTOR & 0.271 & 0.258 & 0.620 & 0.307 & 0.318 & 0.231 & 0.334 \\
& GraphRAG & 0.301 & 0.291 & 0.604 & 0.292 & 0.327 & 0.108 & 0.321 \\
\midrule

\multirow{2}{*}{\textbf{Training-free Agent}} &
Search-o1 & 0.348 & 0.302 & 0.618 & 0.328 & 0.230 & 0.149 & 0.329 \\
& GraphSearch & 0.368 & 0.324 & 0.623 & 0.317 & 0.427 & 0.133 & 0.365 \\
\midrule

\multirow{3}{*}{\textbf{RL-based Agent}} &
Search-R1 & \underline{0.403} & 0.256 & 0.568 & 0.247 & 0.269 & 0.100 & 0.307 \\
& Graph-R1 & 0.267 & 0.274 & 0.464 & 0.298 & 0.407 & \underline{0.289} & 0.333 \\
& Router-R1 & 0.386 & \underline{0.351} & \underline{0.663} & 0.368 & \underline{0.456} & 0.140 & \underline{0.394} \\
\midrule

\multirow{1}{*}{\textbf{Our Method}} &
GraphRAG-Router & \textbf{0.426} & \textbf{0.368} & \textbf{0.672} & \textbf{0.461} & \textbf{0.523} & \textbf{0.443} & \textbf{0.482} \\
\bottomrule
\end{tabular}
\caption{Overall Exact Match (EM) on six QA datasets. $\spadesuit$ denotes the best result among different scale of LLMs. $^\dagger$ denotes the in-domain dataset, and $^\star$ denotes the cross-domain dataset. Bold and underline indicate the best and best baseline results, respectively.}
\label{tab:main}
\vspace{-15pt}
\end{table*}

\section{Experiments}
\subsection{Experimental Setup}
In this section, we conduct extensive experiments to answer the following key research questions (RQs): \textbf{RQ1:} How does GraphRAG-Router perform compared to state-of-the-art baseline models on QA tasks? \textbf{RQ2:} Does GraphRAG-Router yield a better performance-cost trade-off?
\textbf{RQ3:} Can GraphRAG-Router generalize to unseen GraphRAGs and generator LLMs? \textbf{RQ4:} How do different components of the GraphRAG-Router contribute to
performance and cost efficiency?

\noindent\textbf{Datasets.}
We evaluate both GraphRAG-Router and baselines on six QA benchmarks: (1) \textbf{General QA}: Natural Questions (NQ)~\cite{nq}, PopQA~\cite{popqa}, and TriviaQA~\cite{triviaqa}; (2) \textbf{Multi-Hop QA}: HotpotQA~\cite{hotpotqa}, 2WikiMultiHopQA (2Wiki)~\cite{2wiki}, and Musique~\cite{musique}. More details are in Appendix~\ref{app:dataset}.

\noindent\textbf{Baselines.}
We compare GraphRAG-Router with four distinct set of up-to-date, strong baseline models:
(1) \textbf{Basic LLMs}: Direct Inference, Chain-of-Thought (CoT) Prompting~\cite{cot};
(2) \textbf{RAG-based Methods}: Vanilla RAG~\cite{RAG}, GraphRAG~\cite{GraphRAG}, RAPTOR~\cite{RAPTOR}, HippoRAG2~\cite{HippoRAG2}, HyperGraphRAG~\cite{HyperGraph}, and LinearRAG~\cite{LinearRAG};
(3) \textbf{Training-free Agentic Search Systems}: Search-o1~\cite{Search-o1} and GraphSearch~\cite{GraphSearch};
(4) \textbf{RL-based Search Agent}: Search-R1~\cite{Search-R1}, Graph-R1~\cite{Graph-R1}, and Router-R1~\cite{Router-R1}.

% \paragraph{Evaluation Metrics.}
% We evaluate GraphRAGRouter-R1 and baseline models with two different metrics: Exact Match (EM) and F-1.

\noindent\textbf{Implementation Details.}
We pre-train both GraphRAG-Router and baselines on the NQ and HotpotQA datasets. We construct a joint train set of 5K samples from HotpotQA and NQ. For the test, we sample 1000 data points from the original test set or development set. For a fair comparison, we use Qwen2.5-3B-Instruct~\cite{Qwen2.5} as the backbone LLM for all training-based methods. For the SFT stage of our method, we leverage GPT-5.2~\cite{GPT5} to generate 450 general traces and 50 self-reflect traces, respectively. For both RL training stages, we use GRPO~\cite{GRPO} as the default algorithm. For GrapRAGs, we select five representative frameworks: GraphRAG, RAPTOR, HippoRAG2, HyperGraphRAG, and LinearRAG. For generator LLMs, we select 5 cut-edge LLMs that cover three scale ranges: (1) \textbf{Small}: Qwen2.5-7B-Instruct~\cite{Qwen2.5}, LLaMA3.1-8B-Instruct~\cite{llama3}, Ministral-8B-2512~\cite{ministral3}. (2)\textbf{Medium}: Mixtral-8$\times$22B-Instruct~\cite{mixtralexperts} and (3) \textbf{Large}: LLaMA3.3-70B-Instruct~\cite{llama3}. Evaluations are conducted using two metrics, including exact match (EM) and F1-score. More details are in Appendix~\ref{app:implementation}.

\subsection{Overall Performance (RQ1)}
In this section, we evaluate the overall performance of GraphRAG-Router under both in-domain and cross-domain settings. The results are shown in Table~\ref{tab:main}. Based on the results, we observe that:

\textit{\textbf{\underline{Observation 1}: GraphRAG-Router consistently outperforms all baselines on QA tasks.}} As shown in Table~\ref{tab:main}, GraphRAG-Router outperforms competing methods, including basic LLM, RAG-based inference, training-free agent, and RL-based agent, across all six QA benchmarks. While the RL-based agent, such as Router-R1~\cite{Router-R1} improves upon other baselines by interleaved mutlti-turn search and reasoning, GraphRAG-Router achieves even stronger results, especially on multi-hop QA datasets. Notably, GraphRAG-Router exceeds the performance of leading RL-based baselines by an average margin of $+4.71\%$ on general QA and $+38.28\%$ on multi-hop QA, respectively. This highlights the effectiveness of our proposed approach.

\textit{\textbf{\underline{Observation 2}: GraphRAG-Router generalizes well to unseen datasets.}} Despite being pre-trained on NQ and HotpotQA, it maintains strong performance on cross-domain QA datasets, with an average improvement of $+18.55\%$, compared to leading baselines. This indicates that, even pretrained on limited data,  GraphRAG-Router still demonstrates strong generalization capabilities with a transferable routing policy.
\subsection{Can GraphRAG-Router Generalize to Unseen GraphRAGs and LLMs (RQ2)}
To evaluate GraphRAG-Router’s generalization to unseen LLMs and GraphRAG systems, we expand both the routing LLM pool and the GraphRAG pool. Specifically, we introduce two additional state-of-the-art LLMs, Qwen3-8B~\cite{Qwen3} and gpt-oss-120b~\cite{gptoss}, as well as one additional GraphRAG method, LightRAG~\cite{LightRAG}. We then incorporate the descriptions of these newly added LLMs and GraphRAG systems into the routing template. Without any further training, we directly perform inference using the pre-trained GraphRAG-Router. The results are presented in Table~\ref{tab:gen}.
\begin{table}[]
\centering
\footnotesize
\setlength{\tabcolsep}{5pt}
\renewcommand{\arraystretch}{1.3}
\begin{tabular}{lccc}
\toprule
\multirow{2}{*}{\textbf{Model}} & \multicolumn{3}{c}{\textbf{Dataset}} \\ \cline{2-4}
                                & \textbf{NQ} & \textbf{HotpotQA} & \textbf{2Wiki} \\ 
\midrule
\textbf{Router-R1}              & 0.386       & 0.368         & 0.456          \\
\textbf{Router-R1$^*$}          & 0.390       & 0.377         & 0.458          \\
\textbf{GraphRAG-Router}      & 0.426       & \textbf{0.461} & 0.523         \\
\textbf{GraphRAG-Router$^*$}  & \textbf{0.439} & 0.458      & \textbf{0.550} \\
\bottomrule
\end{tabular}
\caption{Results of the generalization capability of GraphRAG-Router.}
\vspace{-4mm}
\label{tab:gen}
\end{table}

\textit{\textbf{\underline{Observation 3}: GraphRAG-Router exhibits strong generalization to previously unseen LLMs and GraphRAG systems, and can be seamlessly extended to new candidates without additional training.}} The results, as shown in Table~\ref{tab:gen}, indicate that GraphRAG-Router achieves comparable or even slightly improved performance across multiple QA datasets. Notably, it establishes new best EM scores on several benchmarks, including NQ and 2Wiki. This suggests that GraphRAG-Router does not merely overfit to the candidate set seen during training but instead learns a robust and transferable routing strategy with strong scalability, enabling it to accommodate newly introduced LLMs and GraphRAG systems effectively.
\subsection{Cost-efficiency Analysis (RQ3)}
\begin{figure*}[t]
    \centering
    \begin{subfigure}[t]{0.49\textwidth}
        \centering
        \includegraphics[width=0.52\linewidth]{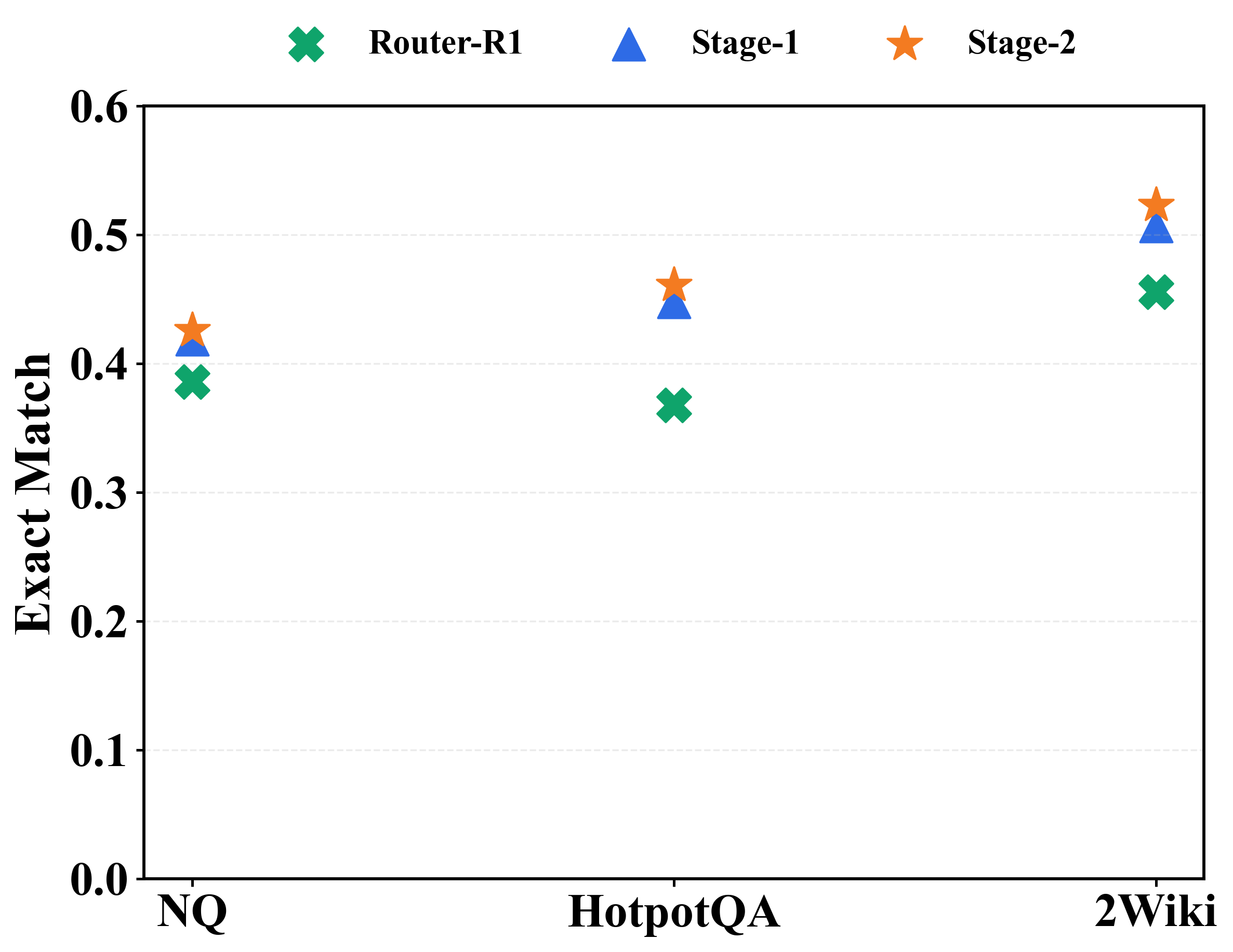}
        \caption{Exact match performance on three datasets.}
        \label{fig:em_eff}
    \end{subfigure}\hspace{-0.8em}
    \begin{subfigure}[t]{0.49\textwidth}
        \centering
        \includegraphics[width=0.52\linewidth]{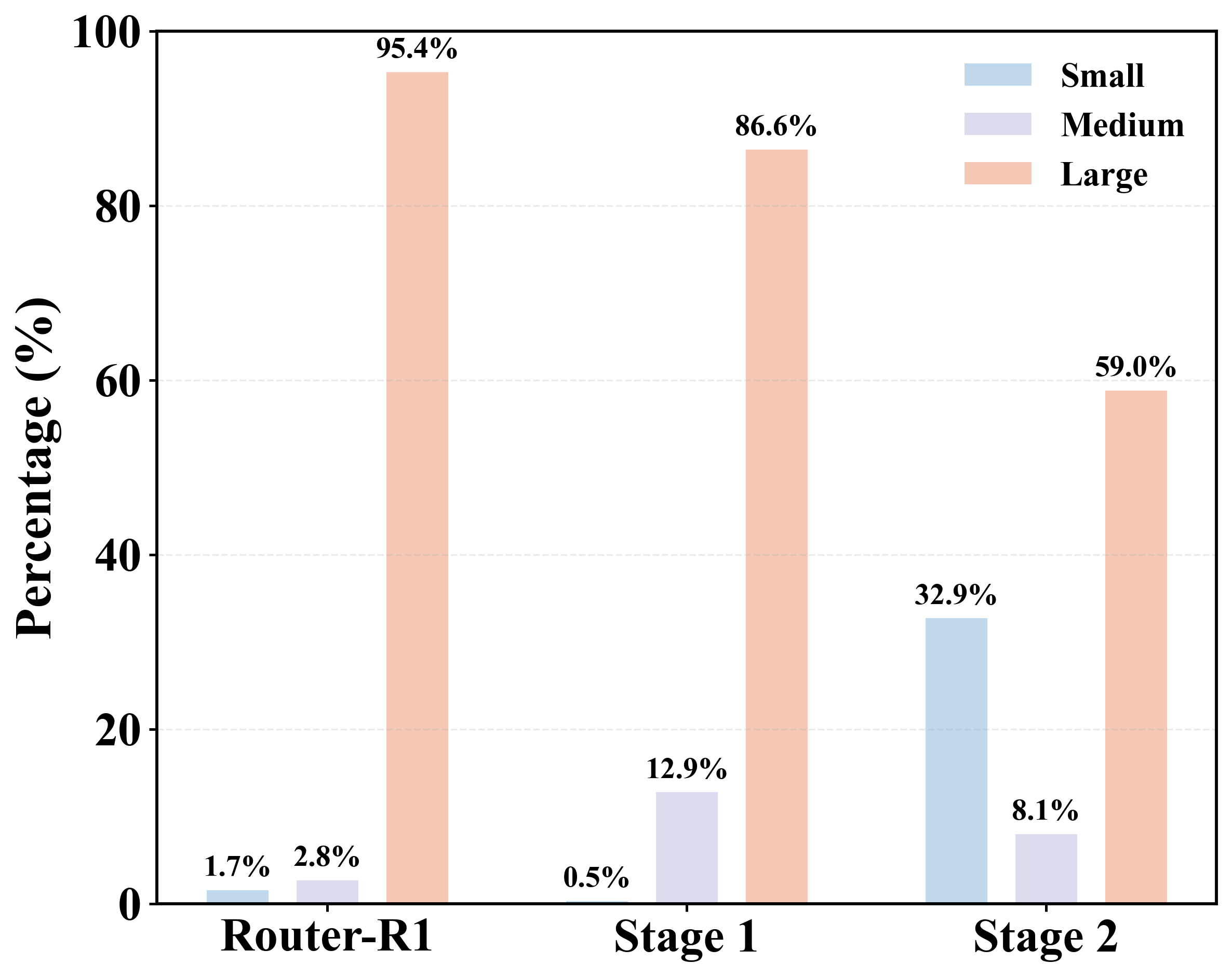}
        \caption{Distribution of routing calls across model scales.}
        \label{fig:count_eff}
    \end{subfigure}
    \caption{Comparison of routing behavior and downstream performance.}
    \label{fig:main_figure}
\end{figure*}

In this section, we analyze the performance-cost trade-off of GraphRAG-Router. Specifically, we present the routing statistics for Stages 1 and 2 across small-, medium-, and large-scale LLMs. The results are shown in Figure~\ref{fig:em_eff} and \ref{fig:count_eff}.

\textit{\textbf{\underline{Observation 4}: GraphRAG-Router achieves a more favorable performance-cost trade-off by reducing reliance on large-scale LLMs.}} In contrast to Stage 1 and Router-R1, which predominantly route queries to large-scale models, GraphRAG-Router reduces large-model usage by nearly $30\%$, reallocating a considerable fraction of queries to medium- and small-scale models. This result highlights the effectiveness of the proposed curriculum cost-aware reward in enabling GraphRAG-Router to adaptively select models that are better aligned with query difficulty while remaining cost-efficient. Moreover, compared with the Stage 1-only setting, GraphRAG-Router still achieves consistent improvement, underscoring its ability to optimize performance while maintaining cost efficiency.
\subsection{Ablation Study (RQ4)}
To answer RQ4, we conduct a comprehensive analysis of each component of GraphRAG-Router.
\subsubsection{Impact of Hierarchical Routing Process}
To further evaluate the effectiveness of the hierarchical routing process, we implement two additional variants of GraphRAG-Router with alternative routing strategies, namely one-time routing and LLM-first routing. For one-time routing, the model directly selects a GraphRAG–LLM pair in a single reasoning process. For LLM-first routing, we reverse the routing order and let the model choose the LLM before selecting the GraphRAG. The results are presented in Table~\ref{tab:routing}, from which we draw the following conclusion:

\textit{\textbf{\underline{Observation 5}: The hierarchical routing strategy is consistently more effective than flat routing alternatives.}} Specifically, compared with one-time routing, hierarchical routing consistently achieves stronger results, showing the advantage of decomposing the routing decision into multiple stages. Moreover, among the hierarchical variants, the GraphRAG-first design outperforms the LLM-first design across all datasets, indicating that deciding the retrieval framework before the generator better aligns with the dependency between evidence acquisition and answer generation.

\begin{table}[]
\centering
\footnotesize
\setlength{\tabcolsep}{5pt}
\renewcommand{\arraystretch}{1.3}
\begin{tabular}{lccc}
\toprule
\multirow{2}{*}{\textbf{Strategy}} & \multicolumn{3}{c}{\textbf{Dataset}} \\ \cline{2-4}
                                   & \textbf{NQ} & \textbf{HotpotQA} & \textbf{2Wiki} \\ 
\midrule
\textbf{One-time}                  & 0.398       & 0.433             & 0.479          \\
\textbf{LLM-First}                 & 0.415       & 0.446             & 0.518          \\
\textbf{GraphRAG-First}            & \textbf{0.426} & \textbf{0.461} & \textbf{0.523} \\
\bottomrule
\end{tabular}
\caption{EM of different routing strategies.}
\vspace{-5pt}
\label{tab:routing}
\end{table}
\subsubsection{Impact of Cold Start SFT}
To assess the effectiveness of SFT, we isolated this stage by training the model from the base model. We then compared their EM and the average number of valid tool calls. From the results in Table~\ref{tab:SFT}, we draw the following conclusion:
\begin{table}[]
\centering
\footnotesize
\setlength{\tabcolsep}{5pt}
\renewcommand{\arraystretch}{1.1}
\begin{tabular}{l|ccc}
\toprule
\textbf{Method} & \textbf{NQ} & \textbf{HotpotQA} & \textbf{Valid call} \\ 
\midrule
w/o SFT         & 0.185       & 0.207         & 0.35                     \\
w/ SFT          & \textbf{0.426} & \textbf{0.461} & \textbf{1.29}         \\
\bottomrule
\end{tabular}
\caption{Ablation study of SFT. Valid call indicates the average valid tool call for both datasets.}
\label{tab:SFT}
\vspace{-10pt}
\end{table}

\textit{\textbf{\underline{Observation 6}: SFT serves to regularize the output format, thereby providing a better foundation for subsequent optimization.}} Removing SFT leads to a significant drop in both EM and the average number of valid tool calls. This indicates that SFT plays a critical role in standardizing the model’s routing behavior.

\subsubsection{Ablation on RL Training Strategies}
To study the impact of our proposed two-stage RL training: Routing Policy Alignment and Difficulty-Aware Cost Optimization, we conduct a stage-wise ablation study. In detail, we ablate each stage and compare the EM of these variants. \textit{\textbf{\underline{Observation 7}: The proposed two-stage RL training significantly enhance the performance of GraphRAG-Router.}} Compare to other variants, GraphRAG-Router with two stage RL training achieves superior performance, with an average improvement of $+2.31\%$. The overall results confirm that both designed training strategies contribute positively to the performance of GraphRAG-Router.
\begin{table}[]
\centering
\footnotesize
\setlength{\tabcolsep}{5pt}
\renewcommand{\arraystretch}{1.3}
\begin{tabular}{ccccc}
\toprule
\multirow{2}{*}{\textbf{SFT}} & \multirow{2}{*}{\textbf{Stage1}} & \multirow{2}{*}{\textbf{Stage2}} & \multicolumn{2}{c}{\textbf{Dataset}} \\ \cline{4-5}
                              &                                   &                                   & \textbf{NQ}    & \textbf{HotpotQA} \\ 
\midrule
\textbf{\cmark}               & \textbf{\xmark}                  & \textbf{\xmark}                  & 0.237          & 0.245        \\
\textbf{\cmark}               & \textbf{\cmark}                  & \textbf{\xmark}                  & 0.419          & 0.448        \\
\textbf{\cmark}               & \textbf{\cmark}                  & \textbf{\cmark}                  & \textbf{0.426} & \textbf{0.461} \\
\bottomrule
\end{tabular}
\caption{Ablation study on training strategies of GraphRAG-Router.}
\label{tab:rl}
\vspace{-10pt}
\end{table}

\section{Conclusion}
In this paper, we introduce GraphRAG-Router, a cost-efficient framework that coordinates GraphRAGs and generator LLMs through agentic routing. GraphRAG-Router leverages a hierarchical routing strategy and a two-stage RL training paradigm to balance performance and inference cost. Across six complex QA benchmarks, it outperforms all state-of-the-art baselines, achieving an average improvement of $22.3\%$ while reducing the overuse of large LLMs by nearly $30\%$.

\section*{Limitation}
While GraphRAG-Router demonstrates strong performance on six QA benchmarks, we acknowledge a few limitations that present opportunities for future work. First, the current study has a limited task scope and primarily considers Wikipedia-based QA settings. it remains an open question whether the proposed GraphRAG-Router can generalize equally well to more complex retrieval-heavy tasks, such as multi-hop reasoning over long enterprise/scientific documents. Second, our current evaluation is conducted mainly in an offline retrieval setting, where the underlying corpora, graph indices, and candidate GraphRAG systems are pre-built and fixed. It remains unclear whether GraphRAG-Router can maintain the same effectiveness in more dynamic online environments with continuously updated knowledge sources and evolving retrieval indices.

% Bibliography entries for the entire Anthology, followed by custom entries
%\bibliography{anthology,custom}
% Custom bibliography entries only
\bibliography{main}

\clearpage
\appendix
% \section{Appendix}
% \label{sec:appendix}
\section{Dataset Details}
\label{app:dataset}
We conduct evaluations on six widely used RAG benchmarks from the FlashRAG toolkit~\cite{FlashRAG}, covering both single-hop and multi-hop question answering tasks:
\begin{itemize}
    \item \textbf{Natural Questions (NQ)~\cite{nq}.} Real user questions from Google Search paired with Wikipedia passages/answers; commonly used for open-domain, mostly single-hop QA.
    \item \textbf{PopQA~\cite{popqa}.} Popular-knowledge question set designed for retrieval-based QA, emphasizing factual queries where the answer must be grounded in retrieved evidence.
    \item \textbf{TriviaQA~\cite{triviaqa}.} Trivia-style questions with evidence documents (often web/Wikipedia); used for open-domain factual QA and long-context evidence matching.
    \item \textbf{HotpotQA~\cite{hotpotqa}.} Multi-hop QA requiring reasoning over multiple supporting Wikipedia passages; includes labeled supporting facts.
    \item \textbf{Musique~\cite{musique}} Multi-hop QA benchmark built to test compositional reasoning across several pieces of evidence, often with more challenging, structured multi-step requirements.
    \item \textbf{2WikiMultiHopQA (2Wiki)~\cite{2wiki}} Multi-hop QA constructed from Wikipedia that typically requires linking two (or more) pages to reach the answer, focusing on cross-article reasoning.
\end{itemize}
\begin{table*}[!h]
\centering
\begin{tabular}{c|c|c|ccc}
\hline
Dataset  & Task         & Knowledge Source & \#Train                                               & \#Dev                                                & \#Test                                           \\ \hline
NQ       & General QA   & Wiki             & 79,168                                                & 8,757                                                & 3,610                                            \\
PopQA    & General QA   & Wiki             & -                                                     & -                                                    & 14,267                                           \\
TriviaQA & General QA   & Wiki \& Web   & 78,785   & 8,837 & 11,313                                           \\
HotpotQA & Multi-hop QA & Wiki             & 90,447                                                & 7,405                                                & -                                                \\
Musique  & Multi-hop QA & Wiki             & 19,938                                                &  2,417 &  - \\
2WikiMultiHopQA    & Multi-hop QA & Wiki             & 15,000                                                & 12,576                                               & -                                                \\ \hline
\end{tabular}
\caption{Dataset Statistics}
\label{tab:dataset}
\end{table*}

\section{Implementation Details}
\label{app:implementation}
\subsection{Baselines}
\paragraph{\textbf{Training-free Search Agents:}}
These approaches do not train an explicit control policy; rather, they use structured prompts and heuristic rules to steer multi-step retrieval and reasoning at inference time. Specifically, we evaluate;
\begin{itemize}
    \item Search-o1~\cite{Search-o1}: Augments large reasoning models with an agentic RAG workflow and a Reason-in-Documents module that refines retrieved evidence before integration, enabling dynamic, noise-reduced knowledge retrieval to improve reliability on complex reasoning tasks and open-domain QA. Our implementation is based on \url{https://github.com/RUC-NLPIR/Search-o1}
    \item GraphSearch~\cite{GraphSearch}: An agentic deep-search workflow for GraphRAG that performs multi-turn, modular retrieval with dual-channel querying over both text chunks (semantic) and structural graphs (relational), consistently improving multi-hop RAG accuracy and generation quality over traditional GraphRAG retrieval. Our implementation is based on \url{https://github.com/DataArcTech/GraphSearch}
\end{itemize}
\paragraph{\textbf{RL-based Search Agent:}}
These approaches use an RL policy to optimize the agent’s search and reasoning behavior, typically adopting GRPO as the reinforcement learning algorithm. Specifically, we choose:
\begin{itemize}
    \item Search-R1~\cite{Search-R1}: A RL-based retrieval-augmented reasoning framework that trains LLMs to autonomously generate multi-turn search queries during step-by-step reasoning, using retrieved-token masking and an outcome-based reward to improve QA performance over standard RAG baselines. Our implementation is based on \url{https://github.com/PeterGriffinJin/Search-R1}
    \item Graph-R1~\cite{Graph-R1}: An agentic GraphRAG framework trained end-to-end with reinforcement learning that builds lightweight knowledge hypergraphs and performs multi-turn retrieval as an agent–environment interaction. Our implementation is based on \url{https://github.com/LHRLAB/Graph-R1}
    \item Router-R1~\cite{Router-R1}: An RL-based multi-LLM routing framework that formulates model selection and response aggregation as a sequential decision-making process. By interleaving internal reasoning with dynamic routing actions, Router-R1 can invoke multiple LLMs adaptively and optimize the trade-off between task performance and inference cost. Our implementation is based on \url{https://github.com/ulab-uiuc/Router-R1}
\end{itemize}

For GraphRAGs in main experiments, we adopt 5 representative GraphRAGs:
\begin{itemize}
    \item \textbf{Hypergraph-based}: HypergraphRAG~\cite{HyperGraph}: A hypergraph-based RAG framework that represents real-world n-ary facts using hyperedges and integrates hypergraph construction, retrieval, and generation. Our implementation is based on \url{https://github.com/LHRLAB/Graph-R1}
    \item \textbf{Entity Graph based}: HippoRAG2~\cite{HippoRAG2}:A memory-inspired RAG framework that extends HippoRAG’s Personalized PageRank retrieval with deeper passage integration and stronger online LLM usage, improving factual, sense-making, and associative memory. Our implementation is based on \url{https://github.com/OSU-NLP-Group/HippoRAG}
    \item \textbf{Tri-Graph based:} LinearRAG~\cite{LinearRAG}: An efficient GraphRAG framework that avoids noisy, costly relation extraction by building a lightweight relation-free hierarchical “Tri-Graph” (via entity extraction + semantic linking) and retrieving evidence with a two-stage process—local entity activation followed by global importance aggregation—yielding stronger and more reliable passage retrieval on multi-hop QA benchmarks. Our implementation is based on \url{https://github.com/DEEP-PolyU/LinearRAG}
    \item \textbf{Tree based}: RAPTOR~\cite{RAPTOR}: A retrieval-augmented approach that builds a hierarchical tree of recursive embeddings, clusters, and bottom-up summaries, enabling inference-time retrieval across long documents at multiple abstraction levels and delivering strong gains. Our implementation is based on \url{https://github.com/parthsarthi03/raptor}
    \item \textbf{Tree based}: GraphRAG~\cite{GraphRAG}:A graph-based QA framework for private corpora that tackles global, corpus-level questions by (1) building an entity knowledge graph and precomputing community summaries, then (2) answering queries via summary-to-partial-response generation followed by a final aggregation, improving comprehensiveness and diversity over standard RAG at million-token scale. our implementation is based on \url{https://microsoft.github.io/graphrag/}
\end{itemize}
For the generalization experiment, we add an additional GraphRAG:
\begin{itemize}
    \item LightRAG~\cite{LightRAG}: A graph-enhanced retrieval-augmented generation framework that integrates graph structures into indexing and retrieval to better capture entity relationships and complex contextual dependencies. By combining graph-based retrieval with vector representations and a dual-level retrieval mechanism, it improves both retrieval accuracy and efficiency. our implementation is based on \url{https://github.com/hkuds/lightrag}
\end{itemize}
For all the GraphRAGs, we utilize the context of each question as the document and organize the corpus by the official settings. For retrieval, we set the top-k as top-5. For the description of each GraphRAG, we adpot GPT-5.2~\cite{GPT5} to summary the key method from the original paper. The detailed description is provided in~\ref{app:template}.
\begin{table*}[!h]
\centering
\begin{tabular}{ccccccc}
\toprule
\textbf{Learning Rate} & \textbf{Batch Size} & \textbf{Epochs} & \textbf{Weight Decay} & \textbf{Optimizer} & \textbf{Lr Scheduler} & \textbf{BF16} \\ \hline
2e-5          & 4          & 2      & 0.01         & Adam      & Cosine       & True \\ \bottomrule
\end{tabular}
\caption{Hyperparameter setting for SFT}
\label{tab:hyperSFT}
\end{table*}

\begin{table*}[!h]
\centering
\large
\renewcommand{\arraystretch}{1.2}
\setlength{\tabcolsep}{12pt}

\begin{tabular}{p{5.5cm} c p{5.5cm} c}
\toprule
\textbf{Hyperparameter} & \textbf{Value} & \textbf{Hyperparameter} & \textbf{Value} \\
\midrule
Learning Rate & 1e-6 & Mini-batch Size & 32  \\
Train Batch Size & 64 & Micro-batch Size & 8 \\
Rollout Group size & 5 & Max Training Steps & 80 \\
KL coefficient & 0.001 & Warm Up Ratio & 0 \\
Max turns & 4 & Max Sequence Length & 4096 \\
Max Response Length & 1024 & Max Length for LLM Response & 600 \\
Tensor Parallel Size & 1 & GPU Utilization Ratio & 0.45 \\
Rollout Temperature (Train) & 1.0 & Rollout Temperature & 1.0 \\
\bottomrule
\end{tabular}
\caption{Hyperparameter settings for stage 1}
\label{tab:hyper_stage1}
\end{table*}

\begin{table*}[!h]
\centering
\large
\renewcommand{\arraystretch}{1.2}
\setlength{\tabcolsep}{12pt}

\begin{tabular}{p{5.5cm} c p{5.5cm} c}
\toprule
\textbf{Hyperparameter} & \textbf{Value} & \textbf{Hyperparameter} & \textbf{Value} \\
\midrule
Learning Rate & 1e-6 & Mini-batch Size & 32  \\
Train Batch Size & 64 & Micro-batch Size & 8 \\
Rollout Group size & 5 & Max Training Steps & 40 \\
KL coefficient & 0.001 & Warm Up Ratio & 0 \\
Max turns & 4 & Max Sequence Length & 4096 \\
Max Response Length & 1024 & Max Length for LLM Response & 600 \\
Tensor Parallel Size & 1 & GPU Utilization Ratio & 0.45 \\
Rollout Temperature (Train) & 1.2 & Rollout Temperature (Eval) & 1.0 \\
\bottomrule
\end{tabular}
\caption{Hyperparameter settings for stage 2}
\label{tab:hyper_stage2}
\end{table*}
\subsection{Generator LLMs}
All the generator LLMs are accessed via OpenRouter APIs\footnote{\url{https://openrouter.ai/}}. We employ the model card as the description of each LLM.
\subsection{GraphRAG-Router}
We use verl\footnote{\url{https://github.com/verl-project/verl}} as our reinforcement learning training framework. The cost of each LLM scale is set to 1 (small), 2 (medium) and 4 (large). The penalty weight of each difficulty level is: 1 (easy), 0.6 (medium), 0.2 (hard). The threshold of $SR(q)$ is set to 0.8. The scaling parameter $\beta$ is set to 0.05. The detailed hyper-parameters of each stage is provided in Table~\ref{tab:hyperSFT}, \ref{tab:hyper_stage1} and \ref{tab:hyper_stage2}.
\begin{table*}[t]
\centering
\small
\setlength{\tabcolsep}{2pt}
\renewcommand{\arraystretch}{1.10}
\label{tab:1}

\begin{tabular}{@{} L{2.8cm} L{3.2cm} *{7}{C{1.35cm}} @{}}
\toprule
\multirow{2}{*}{\textbf{System}} & \multirow{2}{*}{\textbf{Methods}} & \multicolumn{3}{c}{\textbf{General QA}} & \multicolumn{4}{c}{\textbf{Multi-Hop QA}} \\
\cmidrule(r){3-5}\cmidrule(l){6-9}
& & \textbf{NQ$^\dagger$} & \textbf{PopQA$^\star$} & \textbf{TriviaQA$^\star$}
& \textbf{HotpotQA$^\dagger$} & \textbf{2Wiki$^\star$} & \textbf{Musique$^\star$} & \textbf{Avg.} \\
\midrule

\multirow{2}{*}{\textbf{Basic LLM}}
& Direct Infer$^\spadesuit$ & 0.165 & 0.162 & 0.347 & 0.361 & 0.347 & 0.135 & 0.253 \\
& CoT$^\spadesuit$          & 0.399 & 0.316 & 0.618 & 0.436 & 0.344 & 0.243 & 0.393 \\
\midrule

\multirow{6}{*}{\textbf{RAG-Based}}
& Vanilla RAG     & 0.401 & 0.275 & 0.623 & 0.445 & 0.376 & 0.257 & 0.396 \\
& HippoRAG2       & 0.465 & 0.329 & 0.668 & \underline{0.501} & 0.428 & \underline{0.416} & 0.468 \\
& LinearRAG       & 0.436 & 0.334 & 0.661 & 0.487 & 0.469 & 0.403 & 0.465 \\
& HyperGraphRAG   & 0.375 & 0.307 & 0.656 & 0.376 & 0.419 & 0.346 & 0.413 \\
& RAPTOR          & 0.382 & 0.282 & 0.684 & 0.409 & 0.398 & 0.395 & 0.425 \\
& GraphRAG        & 0.423 & 0.348 & 0.705 & 0.388 & 0.401 & 0.248 & 0.419 \\
\midrule

\multirow{2}{*}{\textbf{Training-free Agent}}
& Search-o1    & 0.476 & 0.354 & 0.672 & 0.439 & 0.289 & 0.237 & 0.411 \\
& GraphSearch  & 0.483 & 0.371 & 0.683 & 0.425 & 0.458 & 0.190 & 0.435 \\
\midrule

\multirow{3}{*}{\textbf{RL-based Agent}}
& Search-R1   & 0.473 & 0.357 & 0.617 & 0.371 & 0.451 & 0.160 & 0.405 \\
& Graph-R1    & 0.415 & 0.294 & 0.597 & 0.468 & 0.468 & 0.357 & 0.433 \\
& Router-R1   & \underline{0.495} & \textbf{0.418} & \underline{0.723} & 0.467 & \underline{0.502} & 0.224 & \underline{0.472} \\
\midrule

\multirow{1}{*}{\textbf{Our Method}}
& GraphRAG-Router& \textbf{0.525} & \underline{0.403} & \textbf{0.735} & \textbf{0.589} & \textbf{0.591} & \textbf{0.563} & \textbf{0.568} \\
\bottomrule
\end{tabular}

\caption{Overall F-1 on six QA datasets. $\spadesuit$ denotes the best result among different scale of LLMs. $^\dagger$ denotes the in-domain dataset, and $^\star$ denotes the cross-domain dataset. Bold and underline indicate the best and best baseline results, respectively.}
\label{tab:F1}
\end{table*}
\subsection{Detail for fine-grained format reward}
\label{app:format}
\begin{itemize}[leftmargin=*,itemsep=2pt,topsep=2pt]
    \item \textbf{Fatal format violation} (\(+1.0\)): If the output contains no valid tags, includes unclosed or mismatched tags, or contains nested tags, the format penalty is directly set to \(1.0\).

    \item \textbf{Missing reasoning tag} (\(+0.4\)): A penalty of \(0.4\) is applied if no \texttt{<think>} tag is present.

    \item \textbf{LLM selection before GraphRAG selection} (\(+0.8\)): If a \texttt{<llm>} tag appears before any valid \texttt{<graphrag>} tag, a penalty of \(0.8\) is applied, as this violates the intended hierarchical routing order.

    \item \textbf{Missing GraphRAG selection} (\(+0.4\) / \(+0.6\)): If no valid \texttt{<graphrag>} tag is identified, a penalty of \(0.4\) is applied. If the model proceeds to \texttt{<search>} without a valid GraphRAG selection, the penalty is increased to \(0.6\).

    \item \textbf{Invalid GraphRAG name} (\(+0.2\)): If a \texttt{<graphrag>} tag is present but its content does not correspond to a valid GraphRAG option, a penalty of \(0.2\) is applied.

    \item \textbf{Missing second-stage reasoning} (\(+0.3\)): If a \texttt{<search>} action is produced with fewer than two \texttt{<think>} tags, a penalty of \(0.3\) is applied, encouraging an explicit two-stage reasoning process before execution.

    \item \textbf{Missing LLM selection before search} (\(+0.3\)): If the model issues a \texttt{<search>} action without first specifying a valid \texttt{<llm>} tag, a penalty of \(0.3\) is applied.

    \item \textbf{Invalid LLM name} (\(+0.1\)): If a \texttt{<llm>} tag is present but its content is not a valid LLM name, a penalty of \(0.1\) is applied.

    \item \textbf{Missing search action} (\(+0.4\)): If no \texttt{<search>} tag is present, a penalty of \(0.4\) is applied.

    \item \textbf{Invalid search format} (\(+0.3\)): If the content of \texttt{<search>} does not follow the required format, i.e., it does not contain exactly one ``:'' and one ``;'', a penalty of \(0.3\) is applied.

    \item \textbf{Invalid answer cardinality} (\(+0.3\)): If the output contains zero or more than one \texttt{<answer>} tag, a penalty of \(0.3\) is applied.

    \item \textbf{Empty reasoning content} (\(+0.2\)): If any \texttt{<think>} tag is empty or contains only a placeholder such as ``...'', a penalty of \(0.2\) is applied.
\end{itemize}

The final format penalty is clipped to a maximum value of \(1.0\).
\section{Overall F-1}
The overall F-1 score is shown in Table~\ref{tab:F1}. The results further indicate the effectiveness of our proposed framework.

\section{Additional Experiment Results for SFT}
\begin{figure}[!h]
    \centering
    \includegraphics[width=\linewidth]{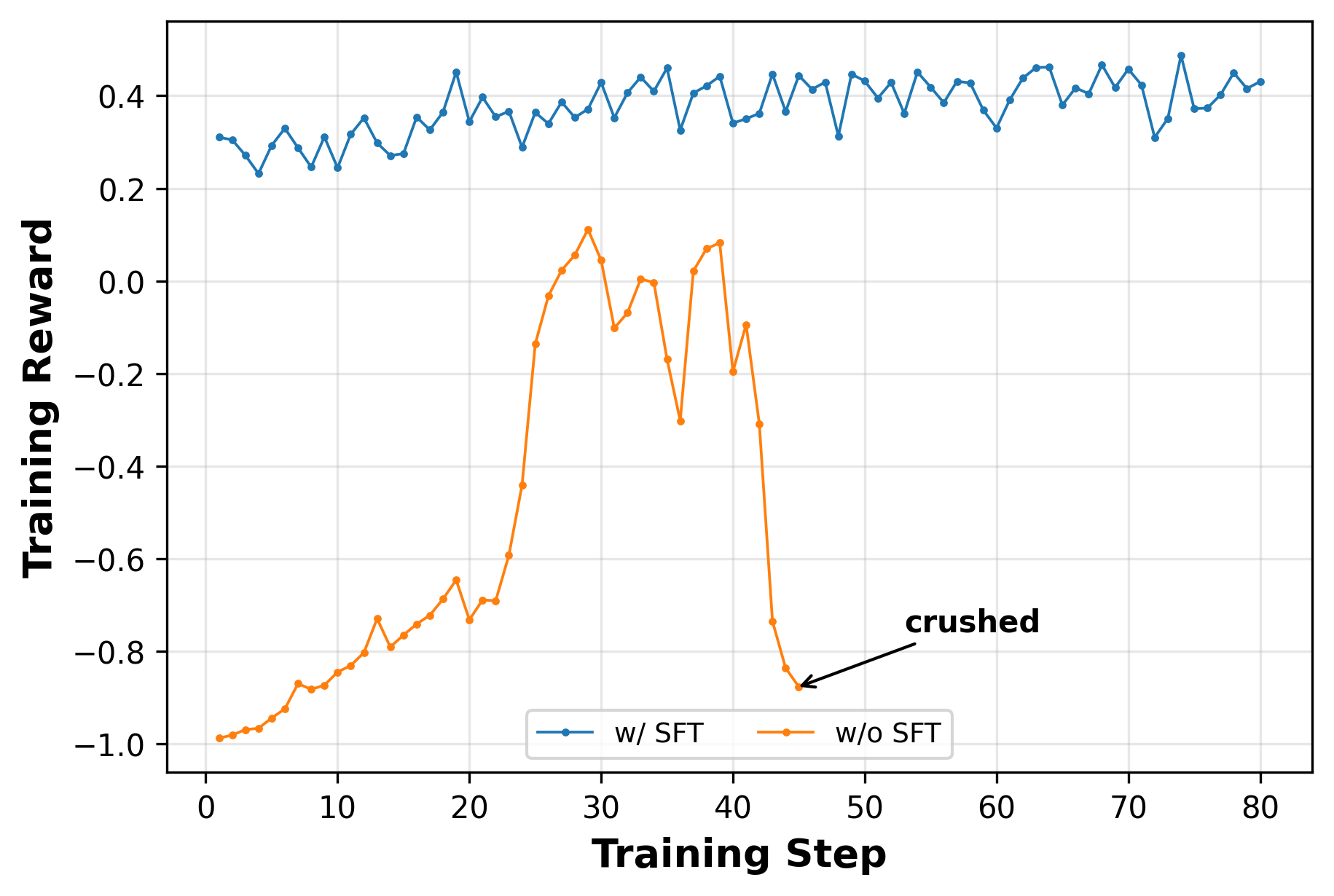}
    \caption{Reward curve}
    \label{fig:reward}
\end{figure}

We provide the reward curves for both variants. As shown in Figure~\ref{fig:reward}, the variant with SFT exhibits a noticeably more stable training process, demonstrating the necessity of SFT for effective optimization.

\section{Ablation on Curriculum Cost-aware Reward}
\begin{figure}[!h]
    \centering
    \includegraphics[width=0.45\textwidth]{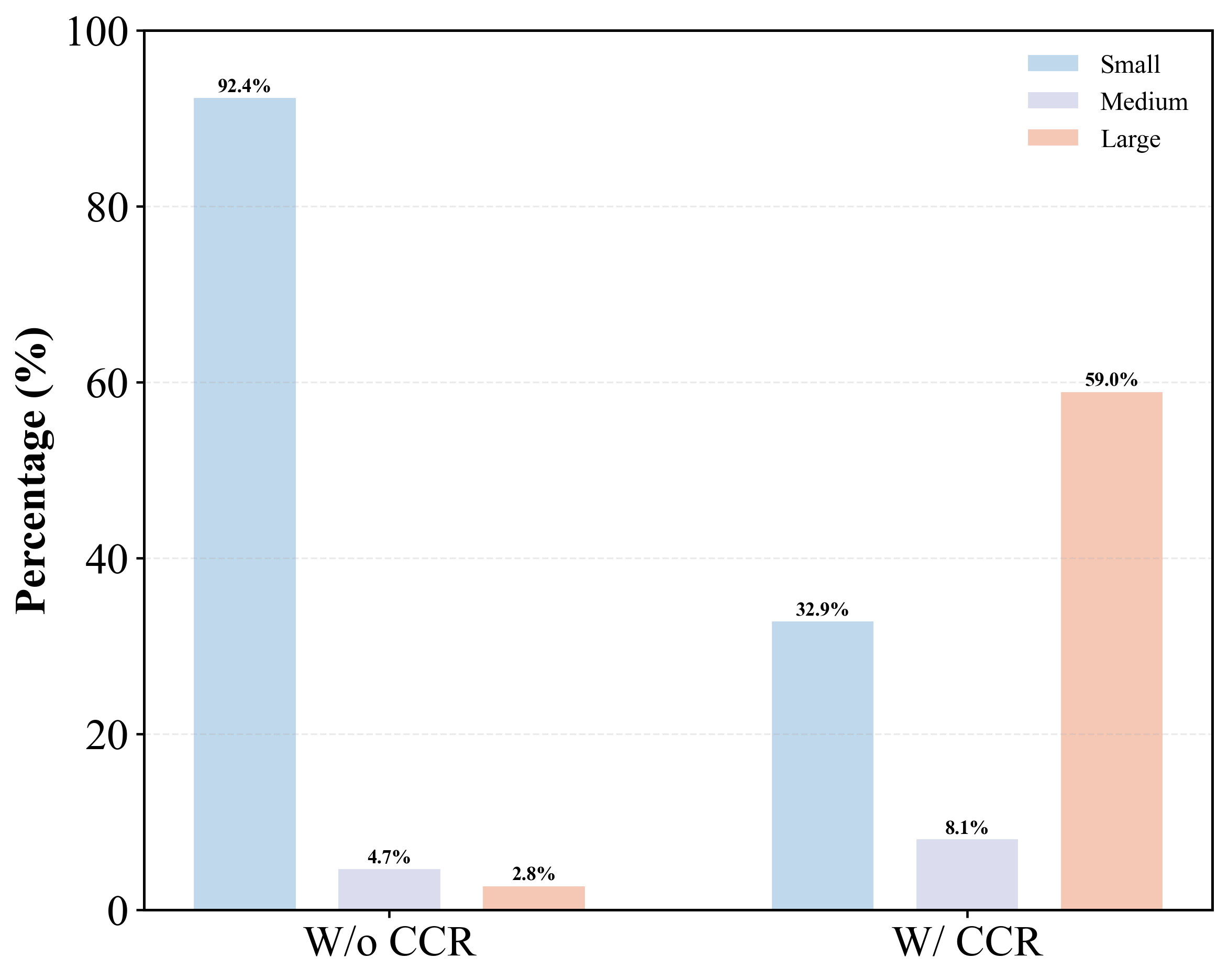}
    \caption{Routing statistics of both reward formulation.}
    \label{fig:curriculum}
\end{figure}

To evaluate the effectiveness of our proposed curriculum cost-aware reward (CCR), we implement a variant that applies the same penalty to all questions, i.e., setting \(W_{\mathrm{Difficulty}(q)} = 1\) for every query. The EM results are reported in Table~\ref{tab:curriculum}, and the routing distribution is shown in Figure~\ref{fig:curriculum}. From the results, we observe that although applying a uniform penalty to all questions can substantially reduce the routing cost, this reduction comes at the expense of answer performance. This suggests that simply imposing a global cost penalty encourages overly conservative routing, whereas our difficulty-aware CCR achieves a better balance between cost efficiency and performance.
\begin{table}[!h]
\centering
\footnotesize
\setlength{\tabcolsep}{5pt}
\renewcommand{\arraystretch}{1.1}
\begin{tabular}{l|cc}
\toprule
\textbf{Method} & \textbf{NQ} & \textbf{HotpotQA} \\ 
\midrule
w/o curriculum         & 0.391      & 0.406 \\
w/ curriculum          & \textbf{0.426} & \textbf{0.461}  \\
\bottomrule
\end{tabular}
\caption{Ablation study of curriculum cost-aware reward.}
\label{tab:curriculum}
\end{table}

\section{Ablation on GraphRAG Routing Pool}
In this section, we conduct an ablation study on the GraphRAG routing pool. Specifically, we consider a GraphRAG-only variant in which the router is restricted to selecting a single predefined GraphRAG framework throughout the entire reasoning process. In other words, the model is not allowed to adaptively switch among multiple GraphRAG candidates, while the generator LLM routing remains unchanged. This setting allows us to examine whether the performance gains of GraphRAG-Router truly come from diverse GraphRAG routing, or whether a single strong GraphRAG is already sufficient. The results in Table~\ref{tab:graph} show that restricting the router to a single GraphRAG consistently underperforms the full routing pool. Although HippoRAG2-only yields the strongest performance among all single-GraphRAG variants, it still falls behind the full model. This suggests that, while certain GraphRAGs can serve as strong default retrievers, no single GraphRAG is sufficient to cover the full diversity of evidence requirements across queries. These results further validate our motivation that no single GraphRAG framework is universally optimal, and that diverse GraphRAG routing is necessary to match the varying evidence requirements of different queries.
\begin{table}[!h]
\centering
\footnotesize
\setlength{\tabcolsep}{5pt}
\renewcommand{\arraystretch}{1.1}
\begin{tabular}{l|cc}
\toprule
\textbf{Method} & \textbf{NQ} & \textbf{HotpotQA} \\ 
\midrule
GraphRAG-only         & 0.393      & 0.408 \\
RAPTOR-only          & 0.376 & 0.417  \\
LinearRAG-only          & 0.398 & 0.433  \\
HyperGraphRAG-only          & 0.368 & 0.401 \\
HippoRAG2-only          & 0.401 & 0.442  \\
ALL          & \textbf{0.426} & \textbf{0.461}  \\
\bottomrule
\end{tabular}
\caption{Ablation study of GraphRAG routing pool.}
\label{tab:graph}
\end{table}

\section{Additional Results for Ablation Study on Training Strategies}
\begin{figure}[!h]
    \centering
    \includegraphics[width=0.45\textwidth]{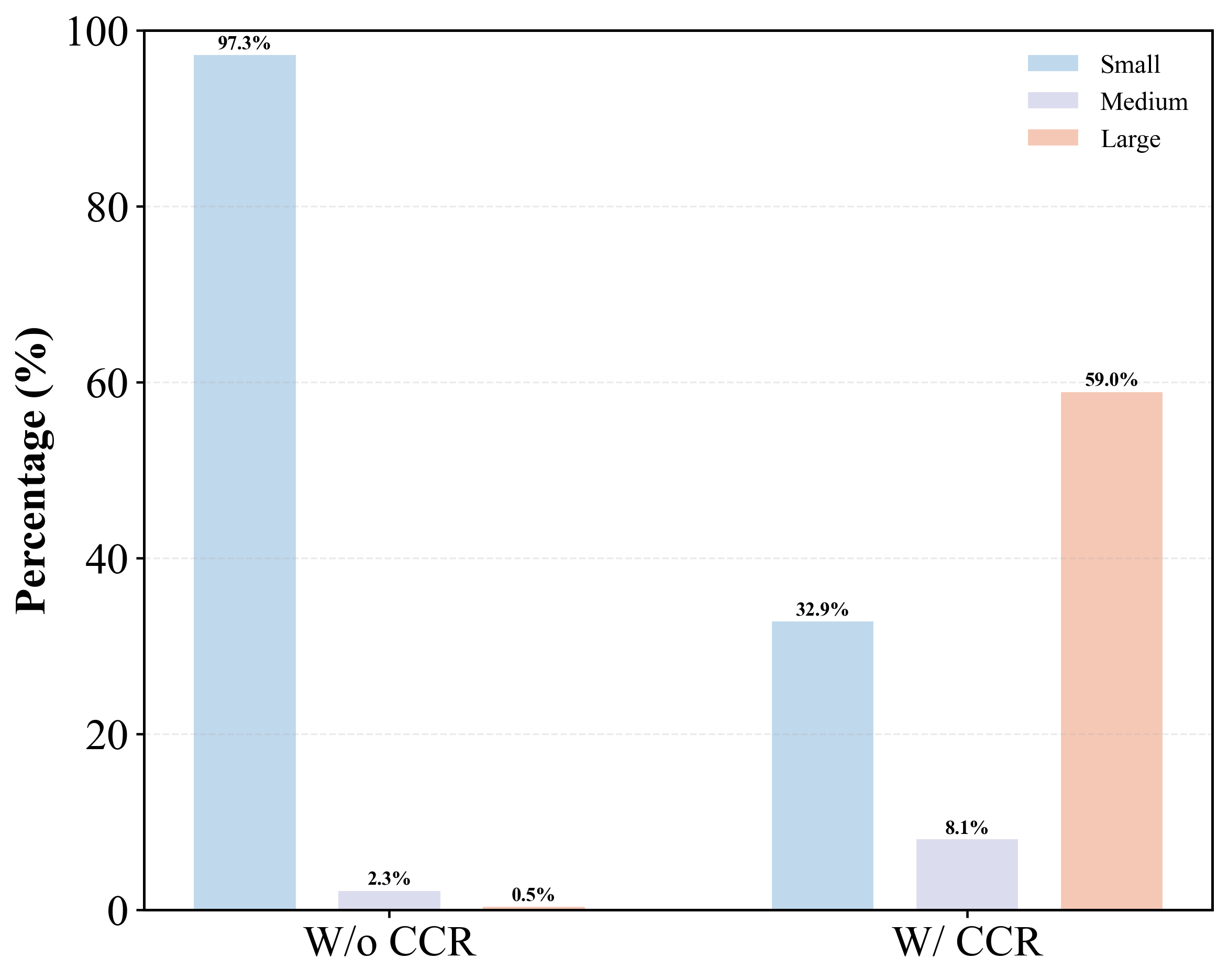}
    \caption{Routing statistics of both RL procedure.}
    \label{fig:RL_procedure}
\end{figure}

In this section, we further investigate the effectiveness of the two-stage RL training procedure. Specifically, we consider a variant in which the two training stages are merged into a single training process. As Table~\ref{tab:RL_procedure} and Figure~\ref{fig:RL_procedure} indicates, this variant exhibits a performance gap compared with the full two-stage training scheme. We attribute this gap to a less favorable performance--cost trade-off during optimization: when effectiveness and cost efficiency are optimized simultaneously from the outset, the model may prematurely favor cheaper routing decisions before learning sufficiently strong routing policies, thereby sacrificing answer quality. This observation indicates the effectiveness of our two-stage RL training paradigm.

\begin{table}[!h]
\centering
\footnotesize
\setlength{\tabcolsep}{5pt}
\renewcommand{\arraystretch}{1.1}
\begin{tabular}{l|cc}
\toprule
\textbf{Method} & \textbf{NQ} & \textbf{HotpotQA} \\ 
\midrule
Mix         & 0.405      & 0.413 \\
Two-stage         & \textbf{0.426} & \textbf{0.461}  \\
\bottomrule
\end{tabular}
\caption{Ablation study of different RL procedure.}
\label{tab:RL_procedure}
\end{table}

\section{Prompt Template}
\label{app:template}

In this section, we describe the prompt templates used in GraphRAG-Router for all experimental settings. We designed a set of task-specific prompts to support different stages of our framework.

\subsection{Instruction of GraphRAG-Router}
We construct a prompt template for GraphRAG-Router, as shown in Figure~\ref{fig:graphrag-router-instruction}. The prompt guides the router through a structured decision process, where it first selects an appropriate external knowledge base and then chooses a generator LLM to process the retrieved information. It also defines a unified tagged format, including \texttt{<graphrag>}, \texttt{<llm>}, \texttt{<search>}, \texttt{<information>}, and \texttt{<answer>}, to make intermediate decisions machine-readable and reduce output ambiguity. In addition, the prompt supports multiple calls to different knowledge bases and LLMs, enabling GraphRAG-Router to leverage the complementary strengths of heterogeneous retrieval systems and language models.

\subsection{Trace Generation Prompt}
\label{app:gen}
We provide prompt templates for both general trace generation and self-reflection trace generation. These templates are designed to produce structured reasoning traces for training GraphRAG-Router. The full prompts are presented in Figure~\ref{fig:generation-prompt} and Figure~\ref{fig:self-reflection-prompt}.

\subsection{GraphRAG Summary Prompt}
Figure~\ref{fig:graphrag-summary-prompt} shows the prompt used to generate concise summaries of candidate GraphRAG frameworks from their original papers. The resulting summaries serve as auxiliary prompt context for routing and trace generation.

\subsection{Descriptions of GraphRAGs and Generator LLMs}
We provide concise descriptions of all candidate GraphRAG frameworks and generator LLMs, as shown in Figure~\ref{fig:llm-descriptions} and Figure~\ref{fig:graphrag-descriptions}. These descriptions summarize the main characteristics and capabilities of each candidate component and are used as auxiliary prompt context for routing and trace generation.

\section{Case Study}
\label{appendix:case_study}

\subsection{Case Study of Cost-efficient Routing}
% In this section, we present a case study of GraphRAG-Router with and without cost-efficient optimization. As shown in Figure~\ref{}

We present a case study of GraphRAG-Router with cost-efficient optimization. As shown in Figure~\ref{fig:case-study1}, the router generates a structured decision trace that explicitly records GraphRAG selection, LLM selection, retrieval results, and final answer generation. This example provides an intuitive illustration of how GraphRAG-Router performs interpretable and evidence-grounded routing.

\subsection{Case Study of Routing Strategies}
Figure~\ref{fig:case-study2} compares the previous one-time routing format with our hierarchical trace format on the same question. The one-time format exhibits two clear limitations, namely inconsistency between reasoning and the actual retrieval action, as well as the inability to revise the answer when the returned information is uncertain or insufficient. In contrast, our format explicitly supports evidence sufficiency checking and second-round rerouting, which enables self-reflection and leads to the correct answer.

\clearpage
\raggedbottom
\onecolumn
% ################################
% Instruction for GraphRAG-Router 
% ################################
% \onecolumn
\begin{tcolorbox}[
    mygreenbox,
    enhanced jigsaw,
    width=\textwidth,
    title={Instruction for GraphRAG-Router},
    fonttitle=\Large\bfseries,
    before upper={
        \setlength{\parindent}{0pt}
        \setlength{\parskip}{0pt}
    }
]
Answer the given question.\par\noindent
Every time you receive new information, you must first conduct reasoning inside \thinktag{<think>} ... \thinktag{</think>}.\par\noindent
If you find you lack some knowledge, you should firstly decide which Knowledge Base to retrieve information from in the base format:
\thinktag{<think>}...\thinktag{</think>}\graphragtag{<graphrag>}GraphRAG-Name\graphragtag{</graphrag>}.\par\noindent
After choosing the Knowledge Base, you must then decide which LLM to use to process the retrieved information in the base format:
\thinktag{<think>}...\thinktag{</think>}\llmtag{<llm>}LLM-Name\llmtag{</llm>}.\par\noindent
When you choose both the Knowledge Base and the specific LLM, you must then retrieve information from the Knowledge Base and use the LLM to process the retrieved information in the base format:
\searchtag{<search>}Question:LLM-Name;GraphRAG-Name\searchtag{</search>}.\par\noindent
The answer from the LLM will be returned between \infotag{<information>} and \infotag{</information>}.\par\noindent
Then you can use the additional information to help you answer the question.

\medskip
!!! STRICT FORMAT RULES for \searchtag{<search>}: !!!\par\noindent
+ You MUST replace LLM-Name with the EXACT name of a model selected from [Qwen2.5-7B-Instruct, LLaMA-3.1-8B-Instruct, LLaMA-3.1-70B-Instruct, Ministral3-8B-2512, Mixtral-8x22B-Instruct].\par\noindent
+ You MUST replace GraphRAG-Name with the EXACT name of a Knowledge Base selected from [HyperGraphRAG, HippoRAG2, RAPTOR, GraphRAG, LinearRAG].\par\noindent
+ You MUST reply in English.\par\noindent
+ You MUST firstly choose the Knowledge Base, then choose the LLM.\par\noindent
+ NEVER copy or paste Model descriptions into \graphragtag{<graphrag>}.\par\noindent
+ NEVER copy or paste GraphRAG descriptions into \llmtag{<llm>}.\par\noindent
+ NEVER output the placeholder format \graphragtag{<graphrag>}GraphRAG-Name\graphragtag{</graphrag>} or \llmtag{<llm>}LLM-Name\llmtag{</llm>}. Always replace all three parts correctly.

\medskip
Before each LLM call and Information Retrieval, you MUST explicitly reason inside \thinktag{<think>} ... \thinktag{</think>} about:\par\noindent
+ Why the specific knowledge base is selected according to the question.\par\noindent
+ Why the specific LLM is selected according to the question and the knowledge base.\par\noindent
+ Which knowledge base is best suited for answering it, based on the GraphRAG's characteristic (described below).\par\noindent
+ Which model is best suited for answering it, based on the LLMs' abilities (described below).

\medskip
When you call an LLM, the response will be returned between \infotag{<information>} and \infotag{</information>}.\par\noindent
You must not limit yourself to repeatedly retrieve from a single knowledge base.\par\noindent
You must not limit yourself to repeatedly calling a single LLM (unless its provided information is consistently the most effective and informative).\par\noindent
You are encouraged to explore and utilize different knowledge bases and LLMs to better understand their respective strengths and weaknesses.\par\noindent
It is acceptable—and recommended—to retrieve from different knowledge bases multiple times for the same input question to gather more comprehensive information. \par\noindent
It is also acceptable—and recommended—to call different LLMs multiple times for the same input question to gather more comprehensive information. \par\noindent

\medskip
\noindent \#\#\#\# The Descriptions of Each LLM: \textbf{\textcolor{myblue}{\{llm\_candidates\_intro\}}}\par\noindent

\medskip
\noindent \#\#\#\# The Descriptions of Each Knowledge Base: \textbf{\textcolor{myblue}{\{graphrag\_candidates\_intro\}}}\par\noindent

\medskip
If you find that no further external knowledge is needed, you can directly provide your final answer inside \answertag{<answer>} ... \answertag{</answer>}, without additional explanation or illustration.

\end{tcolorbox}

\begin{tcolorbox}[
    mygreenbox,
    enhanced jigsaw,
    width=\textwidth,
    notitle,
    before upper={
        \setlength{\parindent}{0pt}
        \setlength{\parskip}{0pt}
    }
]

For example: \answertag{<answer>} Beijing \answertag{</answer>}.  \par\noindent
+ Important: You must not output the placeholder text "\answertag{<answer>} and \answertag{</answer>}" alone.  \par\noindent
+ You must insert your actual answer between \answertag{<answer>} and \answertag{</answer>}, following the correct format.  \par\noindent
Question:\{question\}

\end{tcolorbox}

\captionof{figure}{Instruction for GraphRAG-Router}
\label{fig:graphrag-router-instruction}

% \twocolumn

% ################################
% Generation Prompt 
% ################################
% \onecolumn
\begin{tcolorbox}[
    mygreenbox,
    enhanced jigsaw,
    width=\textwidth,
    title={General Trace Generation Prompt},
    fonttitle=\Large\bfseries,
    before upper={
        \setlength{\parindent}{0pt}
        \setlength{\parskip}{0pt}
    }
]

You are generating structured training examples for fine-tuning a Router model. The Router model is designed to select proper GraphRAG and specialist LLM agent to answer the question.\par\noindent
The Router model uses hierarchical reasoning steps to make decisions: Firstly, it chooses the appropriate GraphRAG based on the question's complexity; Secondly, it selects a specialist LLM agent according to the question and selected GraphRAG.\par\noindent
Your goal is to produce a structured reasoning trace in the specific format shown below and answer the question according to the information provided.\par\noindent

\medskip
You will be given:\par\noindent
+ A user question: \{question\}\par\noindent
+ Short descriptions of available GraphRAGs\par\noindent
+ Short descriptions of available LLM agents\par\noindent
+ A known working pipeline with:\par\noindent
\quad - A GraphRAG model:\{graphrag\}\par\noindent
\quad - A specialist LLM agent:\{llm\}\par\noindent
\quad - Response from the LLM agent:\{information\}\par\noindent

\medskip
Follow these rules strictly:\par\noindent

\medskip
CONTEXT RULES:\par\noindent
+ Output must follow the exact tag order and formatting:\par

{\ttfamily\raggedright\noindent
<think> ... </think>\,\allowbreak
<graphrag>\{graphrag\}</graphrag>\,\allowbreak
<think> ... </think>\,\allowbreak
<llm>\{llm\}</llm>\,\allowbreak
<search>\{question\}:\{llm\};\{graphrag\}</search>\,\allowbreak
<information> ... </information>\,\allowbreak
<think> ... </think>\,\allowbreak
<answer> ... </answer>\par
}

+ DO NOT propose alternative GraphRAGs or LLMs.\par
+ Generate the reasoning trace as you are the Router model selecting the components.\par
+ Generate reasons that are concise and directly relevant to the question.\par
+ The only content inside each \thinktag{<think>} is a short justification phrase or reason for why the given component is appropriate (very brief).\par
+ In the first \thinktag{<think>}, briefly:\par
\quad - analyze what the question is asking for (target entity/field and answer type),\par
\quad - describe the retrieval need,\par
\quad - and state why this matches the selected GraphRAG's strengths.\par
\quad - you should mention the GraphRAG name in the reasoning process.\par
+ In the second \thinktag{<think>}, briefly:\par
\quad  - describe what kind of context the selected GraphRAG is expected to return,\par
\quad  - and state why the selected LLM is a good match to read that context and solve this type of question.\par
\quad  - you should mention the LLM name in the reasoning process.\par
+ In the third \thinktag{<think>}, briefly:\par
\quad  - indicate if you could answer the question based on the information within <information> tags,\par
\quad  - or if an alternative GraphRAG or LLM calls would be needed.\par
+ DO NOT generate factual answers or summaries. \par
+ DO NOT propose alternative GraphRAGs or LLMs. \par
+ DO NOT compare different GraphRAGs or LLMs. \par
+ DO NOT imply ranking, preference, or optimality. \par

\medskip
FORMAT RULES:\par
+ Output MUST follow EXACTLY this tag order with NO extra text:\par

\end{tcolorbox}

\begin{tcolorbox}[
    mygreenbox,
    enhanced jigsaw,
    width=\textwidth,
    notitle,
    before upper={
        \setlength{\parindent}{0pt}
        \setlength{\parskip}{0pt}
    }
]

{\ttfamily\raggedright\noindent
<think>...\allowbreak</think>\,\allowbreak
<graphrag>...\allowbreak</graphrag>\,\allowbreak
<think>...\allowbreak</think>\,\allowbreak
<llm>...\allowbreak</llm>\,\allowbreak
<search>...\allowbreak</search>\,\allowbreak
<information>...\allowbreak</information>\,\allowbreak
<think>...\allowbreak</think>\,\allowbreak
<answer>...\allowbreak</answer>\par
}

+ Do NOT add new tags.\par
+ Do NOT change the tag order.\par
+ Do NOT include any text outside the tags.\par

\medskip
FORBIDDEN LANGUAGE (MUST NOT APPEAR): \par
+ better, best, more suitable, preferred, ideal, optimal \par
+ superior, worse, outperform, more accurate, more efficient \par
+ any form of comparison between GraphRAGs or LLMs \par

\medskip
ALLOWED LANGUAGE (USE ONLY THESE TYPES OF EXPRESSIONS): \par
+ can \par
+ is able to \par
+ supports \par
+ is designed to \par
+ matches the requirement \par
+ is applicable for \par
+ can provide \par

\medskip
Use the following output template:\par

{\ttfamily\raggedright\noindent
<think>Reasons for selecting the given GraphRAG component based on the question.</think>\,\allowbreak
<graphrag>\{graphrag\}</graphrag>\,\allowbreak
<think>Reasons for selecting the given LLM agent component based on the question and selected GraphRAG.</think>\,\allowbreak
<llm>\{llm\}</llm>\,\allowbreak
<search>\{question\}:\{llm\};\{graphrag\}</search>\,\allowbreak
<information>\{information\}</information>\,\allowbreak
<think>Whether the information is sufficient to answer the question or further retrieval/LLM calls are needed.</think>\,\allowbreak
<answer>Your answer</answer>\par
}

\medskip
\noindent \#\#\#\# The Descriptions of Each LLM: \textbf{\textcolor{myblue}{\{llm\_candidates\_intro\}}}\par\noindent

\medskip
\noindent \#\#\#\# The Descriptions of Each Knowledge Base: \textbf{\textcolor{myblue}{\{graphrag\_candidates\_intro\}}}\par\noindent

\end{tcolorbox}

\captionof{figure}{General Trace Generation Prompt}
\label{fig:generation-prompt}

% \twocolumn

% ################################
% Self-Reflection Trace Prompt 
% ################################
% \onecolumn
\begin{tcolorbox}[
    mygreenbox,
    enhanced jigsaw,
    width=\textwidth,
    title={Self-Reflection Trace Generation Prompt},
    fonttitle=\Large\bfseries,
    before upper={
        \setlength{\parindent}{0pt}
        \setlength{\parskip}{0pt}
    }
]

You are generating structured training examples for fine-tuning a Router model.\par\noindent
The Router model is designed to select proper GraphRAG and specialist LLM agent to answer the question.\par\noindent

\medskip
The Router model uses hierarchical reasoning steps to make decisions:\par\noindent
Firstly, it chooses the appropriate GraphRAG based on the question's complexity;\par\noindent
Secondly, it selects a specialist LLM agent according to the question and the selected GraphRAG.\par\noindent

\medskip
Your goal is to produce a structured reasoning trace in the specific format shown below
and answer the question according to the information provided.\par\noindent

\medskip
This task requires generating a TWO-TURN reasoning trace:\par\noindent
\quad - The first turn represents an initial attempt that does NOT provide sufficient information.\par\noindent
\quad - The second turn represents a corrected attempt that IS sufficient to answer the question.\par\noindent

\medskip
Between the two turns, there MUST be exactly ONE \thinktag{<think>} block. \par\noindent
This \thinktag{<think>} block MUST: \par\noindent
\quad - analyze the information from the previous \infotag{<information>} block, and \par\noindent
\quad - explain the decision to retain or change the GraphRAG for the next turn. \par\noindent

\medskip
You will be given: \par\noindent
+ A user question: \{question\} \par\noindent
+ Short descriptions of available GraphRAGs \par\noindent
+ Short descriptions of available LLM agents \par\noindent 
+ A known working pipeline with: \par\noindent
\quad - First-turn GraphRAG model: \{graphrag\_round1\} \par\noindent
\quad - First-turn specialist LLM agent: \{llm\_round1\} \par\noindent
\quad - First-turn response from the LLM agent: \{information\_round1\} \par\noindent
\quad - Second-turn GraphRAG model: \{graphrag\_round2\} \par\noindent
\quad - Second-turn specialist LLM agent: \{llm\_round2\} \par\noindent 
\quad - Second-turn response from the LLM agent: \{information\_round2\} \par\noindent

\medskip
Follow these rules strictly:

\medskip
CONTEXT RULES: \par\noindent
+ Output must follow the exact tag order and formatting shown below. \par\noindent
+ Generate the reasoning trace as if you are the Router model selecting the components. \par\noindent
+ The only content inside each \thinktag{<think>} is a short justification phrase
  explaining applicability or decision rationale. \par\noindent
+ All \thinktag{<think>} blocks MUST be explicitly filled. \par\noindent
+ The first turn MUST conclude that the information is insufficient. \par\noindent
+ The second turn MUST conclude that the information is sufficient. \par\noindent

\medskip
FIRST TURN REASONING: \par\noindent
+ In the first \thinktag{<think>}: \par\noindent
\quad - analyze what the question is asking for (target entity/field and answer type), \par\noindent
\quad - describe the retrieval need, \par\noindent
\quad - explain why the selected GraphRAG (by name) matches this need. \par\noindent
+ In the second \thinktag{<think>}: \par\noindent
\quad - describe the type of context expected from the selected GraphRAG, \par\noindent
\quad - explain why the selected LLM (by name) can process that context. \par\noindent

\end{tcolorbox}

\begin{tcolorbox}[
    mygreenbox,
    enhanced jigsaw,
    width=\textwidth,
    notitle,
    before upper={
        \setlength{\parindent}{0pt}
        \setlength{\parskip}{0pt}
    }
]

BRIDGE THINK (BETWEEN TURNS):\par\noindent
+ The third \thinktag{<think>} MUST:\par\noindent
\quad - analyze whether the information in the first \infotag{<information>} block is sufficient,\par\noindent
\quad - explain why the GraphRAG is retained or changed for the next turn.\par\noindent
+ This \thinktag{<think>} MUST NOT select an LLM.\par\noindent

\medskip
SECOND TURN REASONING:\par\noindent
+ In the fourth \thinktag{<think>}:\par\noindent
\quad - explain why the selected LLM (retained or changed) can process
    the expected context from the second-turn GraphRAG.\par\noindent
+ In the final \thinktag{<think>}:\par\noindent
\quad - clearly state that the information from the second \infotag{<information>}
    is sufficient to answer the question.\par\noindent

\medskip
RESTRICTIONS (APPLY TO ALL TURNS):\par\noindent
+ DO NOT propose alternative GraphRAGs or LLMs beyond those provided.\par\noindent
+ DO NOT compare different GraphRAGs or LLMs.\par\noindent
+ DO NOT imply ranking, preference, or optimality.\par\noindent
+ DO NOT add or remove tags.\par\noindent
+ DO NOT change the tag order.\par\noindent
+ DO NOT include any text outside the tags.\par\noindent

\medskip
FORBIDDEN LANGUAGE (MUST NOT APPEAR):\par\noindent
+ better, best, more suitable, preferred, ideal, optimal\par\noindent
+ superior, worse, outperform, more accurate, more efficient\par\noindent
+ any form of comparison between GraphRAGs or LLMs\par\noindent

\medskip
ALLOWED LANGUAGE (USE ONLY THESE TYPES OF EXPRESSIONS):\par\noindent
+ can\par\noindent
+ is able to\par\noindent
+ supports\par\noindent
+ is designed to\par\noindent
+ matches the requirement\par\noindent
+ is applicable for\par\noindent
+ can provide\par\noindent

\medskip
OUTPUT FORMAT (MUST FOLLOW EXACTLY):\par\noindent
{\ttfamily\raggedright\noindent
% {\raggedright\noindent
<think>...</think>\,\allowbreak
<graphrag>\{graphrag\_round1\}</graphrag>\,\allowbreak
<think>...</think>\,\allowbreak
<llm>\{llm\_round1\}</llm>\,\allowbreak
<search>\{question\}:\{llm\_round1\};\{graphrag\_round1\}</search>\,\allowbreak
<information>\{information\_round1\}</information>\,\allowbreak
<think>...</think>\,\allowbreak
<graphrag>\{graphrag\_round2\}</graphrag>\,\allowbreak
<think>...</think>\,\allowbreak
<llm>\{llm\_round2\}</llm>\,\allowbreak
<search>\{question\}:\{llm\_round2\};\{graphrag\_round2\}</search>\,\allowbreak
<information>\{information\_round2\}</information>\,\allowbreak
<think>...</think>\,\allowbreak
<answer>Your answer</answer>\par
}

\medskip
\noindent \#\#\#\# The Descriptions of Each LLM: \textbf{\textcolor{myblue}{\{llm\_candidates\_intro\}}}\par\noindent

\medskip
\noindent \#\#\#\# The Descriptions of Each Knowledge Base: \textbf{\textcolor{myblue}{\{graphrag\_candidates\_intro\}}}\par\noindent

\end{tcolorbox}
\captionof{figure}{Self-Reflection Trace Generation Prompt}
\label{fig:self-reflection-prompt}
% \twocolumn

% ################################
% GraphRAG Summary Prompt 
% ################################
% \onecolumn
\begin{tcolorbox}[
    mygreenbox,
    enhanced jigsaw,
    width=\textwidth,
    title={GraphRAG Summary Prompt},
    fonttitle=\Large\bfseries,
    before upper={
        \setlength{\parindent}{0pt}
        \setlength{\parskip}{0pt}
    }
]

You are an expert in Graph-based Retrieval-Augmented Generation (RAG) and knowledge graph reasoning.\par\noindent
You will be provided with the full PDF of a research paper about a \textbf{Graph-based Retrieval-Augmented Generation (GraphRAG)} framework.\par\noindent
Read the paper carefully, focusing only on its \textbf{Methodology} (or “Method” / “Approach”) section.\par\noindent
Your task is to summarize the \textbf{Methodology} section of the following research paper about a Graph-based Retrieval-Augmented Generation (GraphRAG) framework. \par\noindent

\medskip
\textbf{Important Constraints}: \par\noindent
    + Use \textbf{only} information explicitly contained in the paper.\par\noindent
    + \textbf{Do not} infer, guess, or add any information beyond what is written.\par\noindent
    + Keep your summary clear, concise, and informative (Limit your response to \textbf{200–300 tokens}). \par\noindent
    + Stay strictly on-topic. Do not include irrelevant or generic content. \par\noindent
    + Do not use bullet points or section headers — output must be \textbf{a single well-formed paragraph}. \par\noindent

\medskip
Focus ONLY on the following aspects: \par\noindent
1. \textbf{Graph Construction Process} \par\noindent
   + How the graph is built (nodes, edges, extraction methods). \par\noindent
   + What embeddings or models are used. \par\noindent

\medskip
2. \textbf{Graph-based Retrieval Mechanism} \par\noindent
   + How information is retrieved from the graph. \par\noindent
   + Granularity (node-level, subgraph-level, etc.) and scoring criteria. \par\noindent

\medskip
3. \textbf{Form of Retrieved Information}  \par\noindent
   + What is passed to the LLM after retrieval (text, triples, summaries, etc.). \par\noindent
   + How it is formatted or integrated for generation. \par\noindent

\medskip
4. \textbf{Distinctive Features}  \par\noindent
    + Unique aspects of this GraphRAG approach. \par\noindent
    + Main innovations or differences from other GraphRAG approaches. \par\noindent

\end{tcolorbox}
\captionof{figure}{GraphRAG Summary Prompt}
\label{fig:graphrag-summary-prompt}

% #####################
% LLM Descriptions 
% #####################
% \onecolumn
\begin{tcolorbox}[
    mygreenbox,
    enhanced jigsaw,
    width=\textwidth,
    title={Descriptions of LLM Candidates},
    fonttitle=\Large\bfseries,
    before upper={
        \setlength{\parindent}{0pt}
        \setlength{\parskip}{0pt}
    }
]

Qwen2.5-7B-Instruct:\par\noindent
Qwen2.5-7B-Instruct is a powerful Chinese-English instruction-tuned large language model designed for tasks in language, coding, mathematics, and reasoning. As part of the Qwen2.5 series, it features enhanced knowledge, stronger coding and math abilities, improved instruction following, better handling of long and structured texts, and supports up to 128K context tokens. It also offers multilingual capabilities across over 29 languages.\par\noindent

\medskip
LLaMA-3.1-8B-Instruct:\par\noindent
LLaMA-3.1-8B-Instruct is an 8-billion-parameter instruction-tuned language model optimized for multilingual dialogue. It provides strong language understanding, reasoning, and text generation performance, outperforming many open-source and closed-source models on standard industry benchmarks.\par\noindent

\medskip
LLaMA-3.1-70B-Instruct:\par\noindent
LLaMA-3.1-70B-Instruct is a 70-billion-parameter state-of-the-art language model designed for advanced multilingual dialogue tasks. It excels in language comprehension, complex reasoning, and high-quality text generation, setting a new standard against both open and closed models in benchmark evaluations.\par\noindent

\medskip
Ministral3-8B-2512:\par\noindent
Ministral3-8B-2512 is a reasoning post-trained version, trained for reasoning tasks, making it ideal for math, coding and stem related use cases. The Ministral 3 family is designed for edge deployment, capable of running on a wide range of hardware. Ministral 3 8B can even be deployed locally, capable of fitting in 24GB of VRAM in BF16, and less than 12GB of RAM/VRAM when quantized.\par\noindent

\medskip
Mixtral-8x22B-Instruct:\par\noindent
Mixtral-8x22B-Instruct is a cutting-edge sparse Mixture-of-Experts (SMoE) large language model from MistralAI. It efficiently uses 39B active parameters out of 141B total, delivering high performance at lower costs. The model excels at following instructions, completing tasks, and generating creative text, with strong skills in multiple languages (English, French, Italian, German, Spanish), mathematics, and coding. It also supports native function calling and handles long contexts up to 64K tokens for better information recall.

\end{tcolorbox}
\captionof{figure}{Descriptions of LLM Candidates}
\label{fig:llm-descriptions}

% #####################
% GraphRAG Descriptions 
% #####################
% \onecolumn
\begin{tcolorbox}[
    mygreenbox,
    enhanced jigsaw,
    width=\textwidth,
    title={Descriptions of GraphRAG Candidates},
    fonttitle=\Large\bfseries,
    before upper={
        \setlength{\parindent}{0pt}
        \setlength{\parskip}{0pt}
    }
]

GraphRAG:\par\noindent
GraphRAG builds a knowledge graph of entities and relationships, then groups them into communities. Retrieval operates at the community level, selecting the most relevant subgraphs and summarizing them for the LLM. Its key strength is global sensemaking—it excels at multi-document, multi-hop reasoning that requires integrating broad context.\par\noindent

\medskip
RAPTOR:\par\noindent
RAPTOR constructs a graph of entities and concepts, represented with embedding models such as BERT. Retrieval combines graph traversal and semantic similarity, returning node-level or subgraph summaries. Its main advantage is capturing both semantic context and graph structure, making it strong for relational reasoning. \par\noindent

\medskip
HippoRAG2:\par\noindent
HippoRAG2 builds a knowledge graph using OpenIE to extract entities, passages, and triples. It retrieves information using Personalized PageRank, focusing on the most relevant nodes and triples. Its strength lies in clear, structured factual retrieval, making it ideal for queries that rely on explicit fact chains. \par\noindent

\medskip
HyperGraphRAG:\par\noindent
HyperGraphRAG uses hyperedges to represent n-ary relationships involving multiple entities, not just binary relations. Retrieval scores both entities and hyperedges using semantic similarity to find the most relevant high-order facts. Its unique strength is expressing complex multi-entity relationships, making it suitable for advanced reasoning tasks. \par\noindent

\medskip
LinearRAG:\par\noindent
LinearRAG builds a lightweight Tri-Graph of entities, sentences, and passages using simple NER rather than heavy relation extraction. Its retrieval combines local entity activation with global PageRank to gather highly relevant contextual passages. The main advantage is low noise, high scalability, and fast retrieval, ideal for large-scale or real-time scenarios. \par\noindent

\end{tcolorbox}
\captionof{figure}{Descriptions of GraphRAG Candidates}
\label{fig:graphrag-descriptions}

\clearpage
\twocolumn
\flushbottom

% ======================================================

% ========================
% Case Study 1
% ========================
\clearpage
% \section{Case Study}
% \label{appendix:case_study}

\begin{center}
\begin{minipage}{0.96\textwidth}

\begin{tcolorbox}[
  casestudybox,
  width=\linewidth,
  title={\textbf{Question:} What is bermuda competing in the winter olympics?\\[2mm]
  \textbf{Ground Truth:} Cross-country skiing}
]
Response of \textbf{GraphRAG-Router}: \par\medskip

\thinktag{<think>}The question asks for the sport Bermuda is competing in during the Winter Olympics; HippoRAG2 supports structured retrieval of explicit fact chains via nodes and triples.\thinktag{</think>} \par

\graphragtag{<graphrag>}HippoRAG2\graphragtag{</graphrag>}\par

\thinktag{<think>}HippoRAG2 can provide passages/triples that link Bermuda to the relevant Winter Olympics event; Qwen2.5-7B-Instruct is able to read that retrieved context and produce the requested single-entity answer.\thinktag{</think>} \par

\llmtag{<llm>}Qwen2.5-7B-Instruct\llmtag{</llm>} \par

% \searchtag{<search>}what is bermuda competing in the winter olympics?:Qwen2.5-7B-Instruct;HippoRAG2\searchtag{</search>} \par
{\raggedright
\searchtag{<search>}what is bermuda competing in the winter olympics?\allowbreak:\allowbreak Qwen2.5-7B-Instruct\allowbreak;\allowbreak HippoRAG2\searchtag{</search>}\par
}

\infotag{<information>}Answer: Cross-country skiing. Explanation: In the 2018 Winter Olympics, Bermuda sent one athlete, cross-country skier Tucker Murphy, to compete in the 15 km freestyle event. This continued a trend of Bermuda being represented solely by a cross-country skier in the three consecutive Winter Olympics up until 2018.\thinktag{</information>} \par

\thinktag{<think>}The information includes the needed statement directly, so it is sufficient to answer without further retrieval or additional LLM calls.\thinktag{</think>} \par

\answertag{<answer>}Cross-country skiing\answertag{</answer>}

\end{tcolorbox}

\captionsetup{justification=centering,singlelinecheck=true}
\captionof{figure}{Case study of GraphRAG-Router under cost-efficient routing}
\label{fig:case-study1}

\end{minipage}
\end{center}

% =========================================
% Case Study 2
% =========================================
\clearpage
\onecolumn

% % ---------- colors ----------
\definecolor{caseblue}{RGB}{31,78,121}
\tcbset{
  schemestrip/.style={
    enhanced,
    breakable=false,
    width=\linewidth,
    colback=blue!9,
    colframe=blue!9,
    boxrule=0pt,
    borderline west={1.2pt}{0pt}{caseblue},
    arc=0mm,
    outer arc=0mm,
    left=3mm,
    right=1mm,
    top=1mm,
    bottom=1mm,
    boxsep=0pt
  }
}

\begin{tcolorbox}[
  casestudybox,
  width=\linewidth,
  title={\textbf{Question:} When did the the Antonine who was known as the third of the five good emperors live?\\[2mm]
  \textbf{Ground Truth:} 24 January 76 – 10 July 138}
]

% -----------------------------------------
% Scheme 1
% -----------------------------------------
\begin{tcolorbox}[schemestrip]
\textbf{One-time routing (incorrect)}
\end{tcolorbox}

Response of \textbf{GraphRAG-Router}: \par\medskip

\thinktag{<think>}I don't have enough information to answer that question, but I do know that the Antonine dynasty is an historical topic so a specialized knowledge base is necessary. The RAPTOR knowledge base is well-suited for retrieval of structured facts, allowing me to understand when historical person were alive. A model that specializes in instruction-tuning may also perform well as these models often feature instruction following and provide strong language-based reasoning.\thinktag{</think>} \par

{\raggedright
\searchtag{<search>}LLaMA-3.1-70B-Instruct: When did the the Antonine who was known as the third of the five good emperors live?;HippoRAG2\searchtag{</search>}\par
}

\infotag{<information>}A nice historical question! According to the context, I don't see any direct information about the Antonine emperors. However, I can provide some helpful background knowledge. The ``Five Good Emperors'' is a term used to describe a period of relative peace and stability in the Roman Empire, which occurred during the reigns of five consecutive emperors: Nerva, Trajan, Hadrian, Antoninus Pius, and Marcus Aurelius. The Antonine in question is likely Antoninus Pius, who was the fourth emperor of the Nerva-Antonine dynasty. He ruled from 138 AD to 161 AD. So, to answer your question, Antoninus Pius, the third of the five good emperors, lived from approximately 86 AD to 161 AD. Please let me know if this information is helpful, or if I can assist further!\infotag{</information>} \par

\answertag{<answer>}138 AD to 161 AD\answertag{</answer>}

\medskip

% -----------------------------------------
% Scheme 2
% -----------------------------------------
\begin{tcolorbox}[schemestrip]
\textbf{Hierarchical routing (correct)}
\end{tcolorbox}

Response of \textbf{GraphRAG-Router}: \par\medskip

\thinktag{<think>}Question asks for the birth/death date of the Antonine known as the third of the five good emperors; needs explicit factual retrieval of a short entity timeline, which HippoRAG2 supports.\thinktag{</think>} \par

\graphragtag{<graphrag>}HippoRAG2\graphragtag{</graphrag>} \par

\thinktag{<think>}HippoRAG2 can provide a concise factual context via structured triples linking name and lifespan; LLaMA-3.1-70B-Instruct is able to read that context and produce the requested single-date answer.\thinktag{</think>} \par

\llmtag{<llm>}LLaMA-3.1-70B-Instruct\llmtag{</llm>} \par

{\raggedright
\searchtag{<search>}When did the the Antonine who was known as the third of the five good emperors live?: LLaMA-3.1-70B-Instruct; HippoRAG2\searchtag{</search>}\par
}

\infotag{<information>}Explanation: I cannot confidently answer this question based on the given Context. The Context mentions that Hadrian is considered by some to have been a humanist, and he is regarded as the third of the Five Good Emperors. However, it does not explicitly state that Antoninus Pius is the third of the Five Good Emperors. Instead, it mentions that Antoninus Pius was one of the Five Good Emperors, but it does not provide his ranking. Please consult other specialized LLMs or retrieve from other knowledge bases for further assistance.\infotag{</information>} \par

\thinktag{<think>}The first information block states it cannot answer based on the given context, so it is insufficient; switch to RAPTOR to support relational retrieval connecting the name to an explicit date.\thinktag{</think>} \par

\graphragtag{<graphrag>}RAPTOR\graphragtag{</graphrag>} \par

\thinktag{<think>}LLaMA-3.1-70B-Instruct is able to use RAPTOR’s factual retrieval snippets to locate an explicit statement about when Antoninus Pius lived and is able to provide the requested date succinctly.\thinktag{</think>} \par

\llmtag{<llm>}LLaMA-3.1-70B-Instruct\llmtag{</llm>} \par

{\raggedright
\searchtag{<search>}When did the the Antonine who was known as the third of the five good emperors live?: LLaMA-3.1-70B-Instruct; RAPTOR\searchtag{</search>}\par
}

\infotag{<information>}Answer: 24 January 76 -- 10 July 138. Explanation: According to the Context, Publius Aelius Hadrianus Augustus, also known as Hadrian, was the Roman emperor from 117 to 138 and is considered the third of the Five Good Emperors.\infotag{</information>} \par

\thinktag{<think>}The second information block states the explicit lifespan for Antoninus Pius (Hadrian), so it is sufficient to answer the question.\thinktag{</think>} \par

\answertag{<answer>}24 January 76 -- 10 July 138\answertag{</answer>}

\end{tcolorbox}

\captionsetup{justification=centering,singlelinecheck=true}
\captionof{figure}{Comparison between one-time routing and hierarchical routing on the same question}
\label{fig:case-study2}

\clearpage
\twocolumn

\end{document}